\documentclass[conference]{IEEEtran}
\IEEEoverridecommandlockouts
\ifCLASSINFOpdf
\else
\fi
\usepackage{graphicx}
\usepackage{subfigure}

\usepackage{amsthm, mathtools, bm, amsmath, amssymb, arydshln}
\usepackage{bbm,boondox-cal}
\usepackage{multirow, multicol}
\usepackage[lined,boxed,commentsnumbered, ruled, noend]{algorithm2e}

\newtheorem{LL}{Lemma}
\newtheorem{TT}{Theorem}
\newtheorem{CC}{Corollary}

\newtheorem{DD}{Definition}

\newtheorem{RM}{Remark}

\usepackage{color}

\newcommand{\chr}{\color{black}}  %INFOCOM to ICDCS revise
\newcommand{\tabincell}[2]{\begin{tabular}{@{}#1@{}}#2\end{tabular}}

\newcommand{\secref}[1]{Sec.~\ref{#1}}

\newcommand{\algref}[1]{Algorithm~\ref{#1}}
\newcommand{\figref}[1]{Fig.~\ref{#1}}
\newcommand{\tabref}[1]{Table~\ref{#1}}
\newcommand{\ttref}[1]{Theorem~\ref{#1}}
\newcommand{\llref}[1]{Lemma~\ref{#1}}
\newcommand{\ccref}[1]{Corollary~\ref{#1}}
\newcommand{\ddref}[1]{Definition~\ref{#1}}
\newcommand{\equref}[1]{Eq.~(\ref{#1})}

\usepackage{cite, url}
\begin{document}
	
	\title{Joint Storage Allocation and Computation Design for Private Edge Computing}
   %\title{A Fundamental Tradeoff between Storage and Communication in Private Coded Edge Computing}

\author{\IEEEauthorblockN{Jiqing Chang$^{\dagger}$, Jin Wang$^{\dagger }$, Kejie Lu$^\star$, Lingzhi Li$^{\dagger }$, Fei Gu$^{\dagger }$, Jianping Wang$^\ddagger$
			%$^\#$
			%\thanks{$^*$Corresponding author: Jin Wang, wjin1985@suda.edu.cn.}
			}
			\IEEEauthorblockA{\small
			$^\dagger$ School of Computer Science and Technology, Soochow University\\
$^\dagger$ Collaborative Innovation Center of Novel Software Technology and Industrialization\\
			$^\star$ Department of Computer Science and Engineering, University of Puerto Rico at Mayag\"{u}ez\\
$^\ddagger$ Department of Computer Science, City University of Hong Kong\\
}}

	\maketitle

\begin{abstract}
	In recent years, edge computing (EC) has attracted great attention for its high-speed computing and low-latency characteristics.
However, there are many challenges in the implementation of EC. Firstly, user's privacy has been raised as a major concern because the edge devices may be untrustworthy. In the case of \emph{Private Edge Computing} (PEC), a user wants to compute a matrix multiplication between its local matrix and one of the matrices in a library, which has been redundantly stored in edge devices. When utilizing resources of edge devices, the privacy requires that each edge device cannot know which matrix stored on it is desired by the user for the multiplication. Secondly, edge devices usually have limited communication and storage resources, which makes it impossible for them to store all matrices in the library.
In this paper, we consider the limited resources of edge devices and propose an unified framework for PEC. Within the framework, we study two highly-coupled problems, (1) storage allocation, that determines which matrices are stored on each edge device, and (2) computation design, that determines which matrices (or linear combinations of them) in each edge device are selected to participate in the computing process with the privacy consideration.
Specifically, we give a general storage allocation scheme and then design two feasible private computation schemes, {\em i.e.}, {\em General Private Computation} (GPC) scheme and \emph{Private Coded Computation} (PCC) scheme. In particular, GPC can be applied in general case and PCC can only be applied in special cases, while PCC achieves less communication load.
We theoretically analyze the proposed computing schemes and compare them with other schemes. Finally, we conduct extensive simulations to show the effectiveness of the proposed schemes.
%We also theoretically analyze the proposed schemes and show that they enable a tradeoff between the storage and communication resources.
	\end{abstract}

		
	\pagestyle{plain}
	\thispagestyle{plain}
	
	\section{Introduction}
	\label{introduction}
	Recently, {\em edge computing} (EC) has attracted great attention because it extends the service resources of cloud to the edge of network \cite{CDC2}.
EC can provide high-speed computing and low-latency service to support many distributed computing tasks \cite{C3,C15,C14,C9} and latency-sensitive applications, e.g., Internet-of-Things (IoT), data analytics, crowdsourcing, and virtual/augmented/mixed reality (VR/AR/MR) \cite{C1}.

	Although EC has been shown as a promising technique for distributed computing, it faces many challenges. Firstly, edge devices usually have limited computation, communication and storage resources. Secondly, distributing tasks to untrustworthy edge devices raises major concerns about the security problem, including confidentiality of information and privacy of user. To address both challenges, \emph{coded distributed computing} (CDC) has been proposed and applied to perform different distributed computing tasks with lower calculation delay and communication load \cite{Lisongze4,CDC1,M3, M2,M8, M1,M9,M5 }.

However, for the security problem of CDC, the confidentiality of information has been widely studied \cite{C15,C14,C9,M6,Cao2019,Zhou2019}, while the privacy of user has not been fully investigated. In particular, for the privacy problem in CDC, most related works focus on the computation design and capacity analysis \cite{M1, M2, M3, M5,M8,M9}. For example, \cite{M5,M9} considered the \emph{private information retrieval} (PIR) problem and given the information theoretic capacity, which is the maximum number of bits of desired information that can be privately retrieved per bit of downloaded information.
The \emph{private function retrieval} (PFR) have been studied in \cite{M8, M2, M3}. The authors proposed computing schemes to compute an arbitrary function of database, while keeping the function hidden by combining it with other data.
In these studies, the computing devices either store the whole library or download the data from the public library for each computing task. They are not suitable for EC because edge devices usually have limited storage and communication resources.

In this paper, we consider the private computation in EC scenario, where each edge device may only store part of the library. Specifically, we take matrix multiplication \cite{M1,C15,C17,C14,M6}, which is a critical and indispensable operation underlying many distributed machine learning algorithms \cite{MM1}, as a representative computing task. We will study two highly-coupled problems, (1) storage allocation, that determines which matrices are stored on each edge device, and (2) computation design, that determines which matrices (or linear combinations of them) in each edge device are selected to participate in the computing process with the privacy consideration.

\begin{figure*}[htb] \centering
\begin{minipage}[t]{0.2\linewidth}
\centering
\includegraphics[width=1.5in]{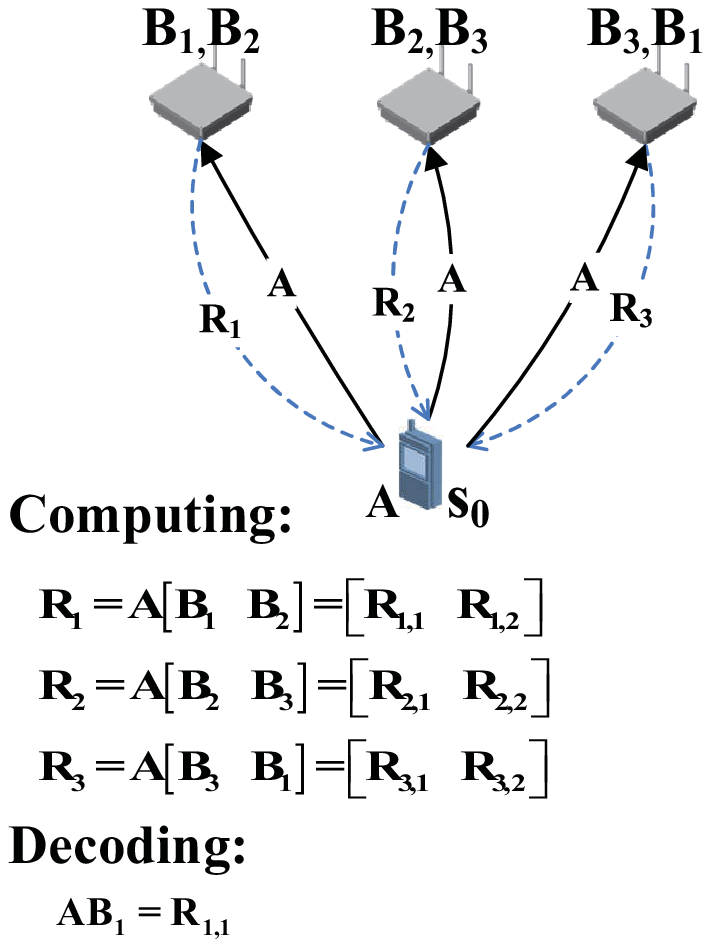}

\mbox{\footnotesize (a) RPC}
\end{minipage}%
\begin{minipage}[t]{0.28\linewidth}
\centering
\includegraphics[width=1.7in]{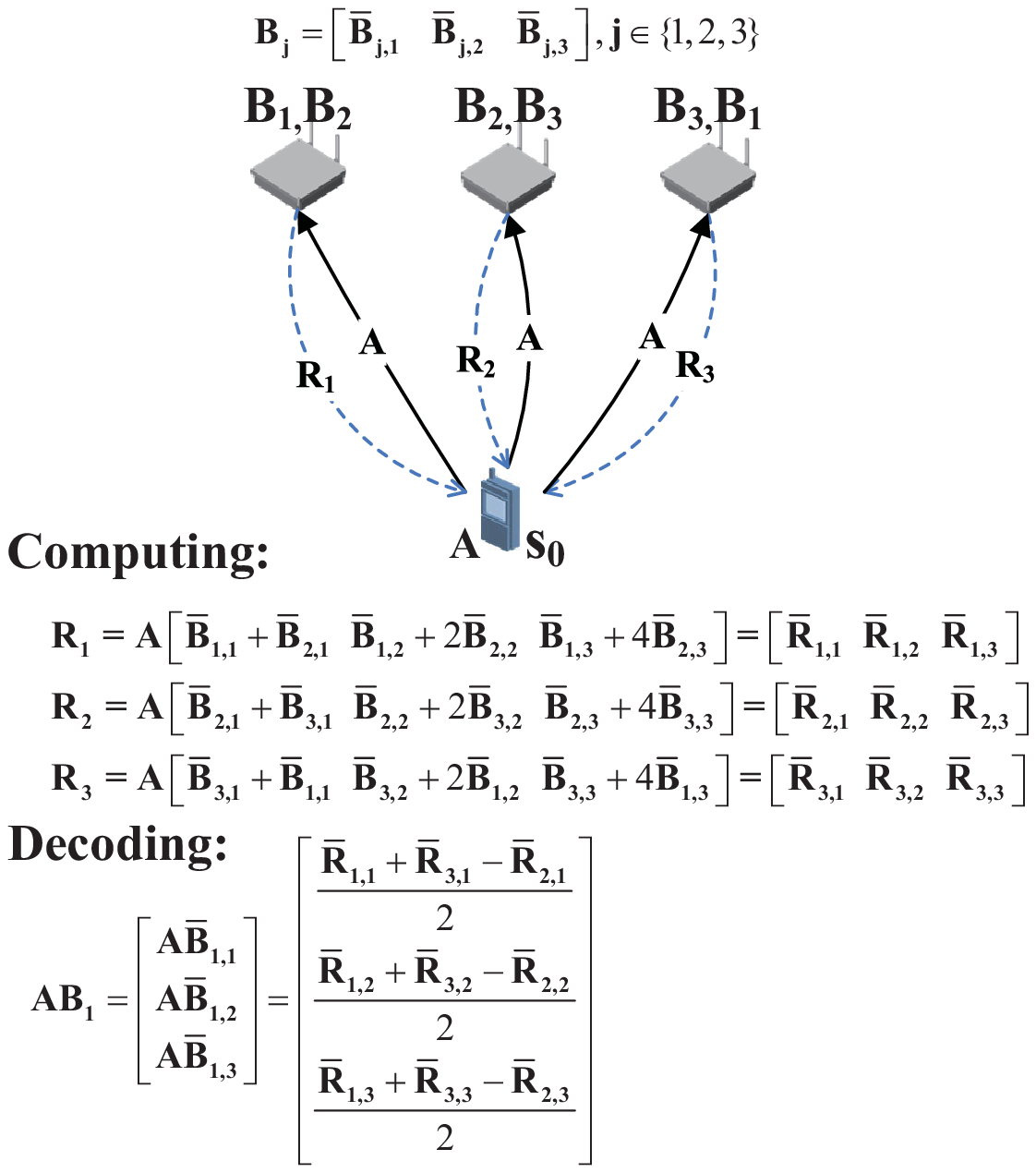}%P4.eps}

\mbox{\footnotesize (b) VPC}
%\mbox{\footnotesize  private computing scheme.}
\end{minipage}
\begin{minipage}[t]{0.2\linewidth}
\centering
\includegraphics[width=1.45in]{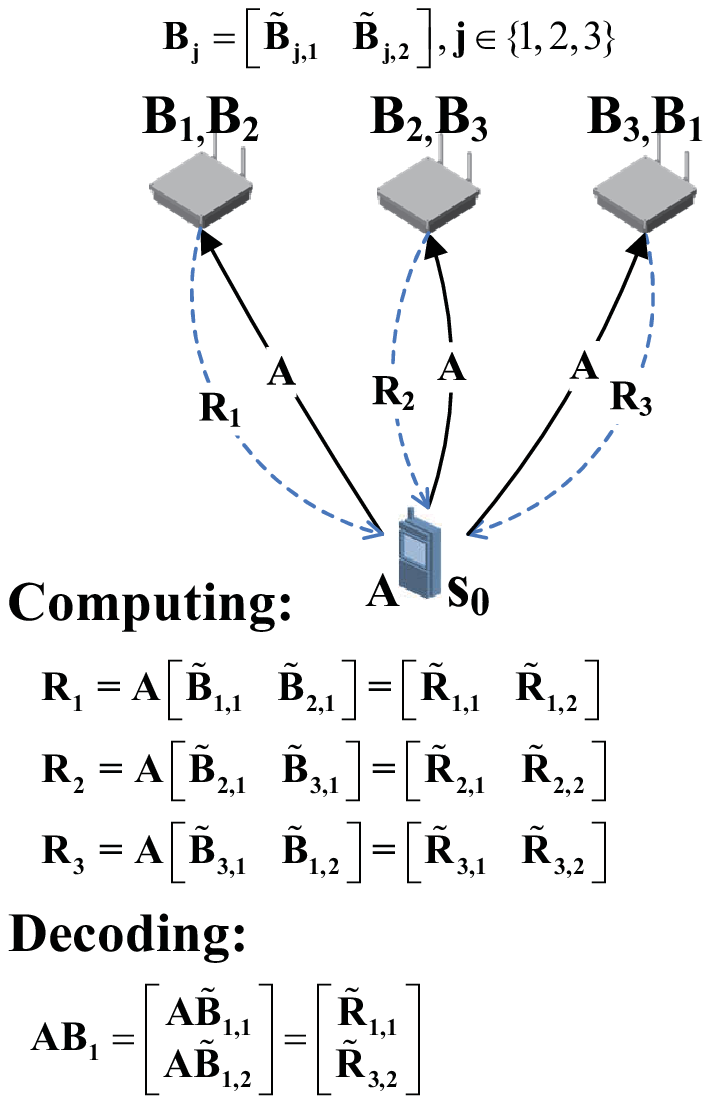}

\mbox{\footnotesize (c) GPC}
%\mbox{\footnotesize  private computing scheme.}
\end{minipage}
\begin{minipage}[t]{0.25\linewidth}
\centering
\includegraphics[width=1.65in]{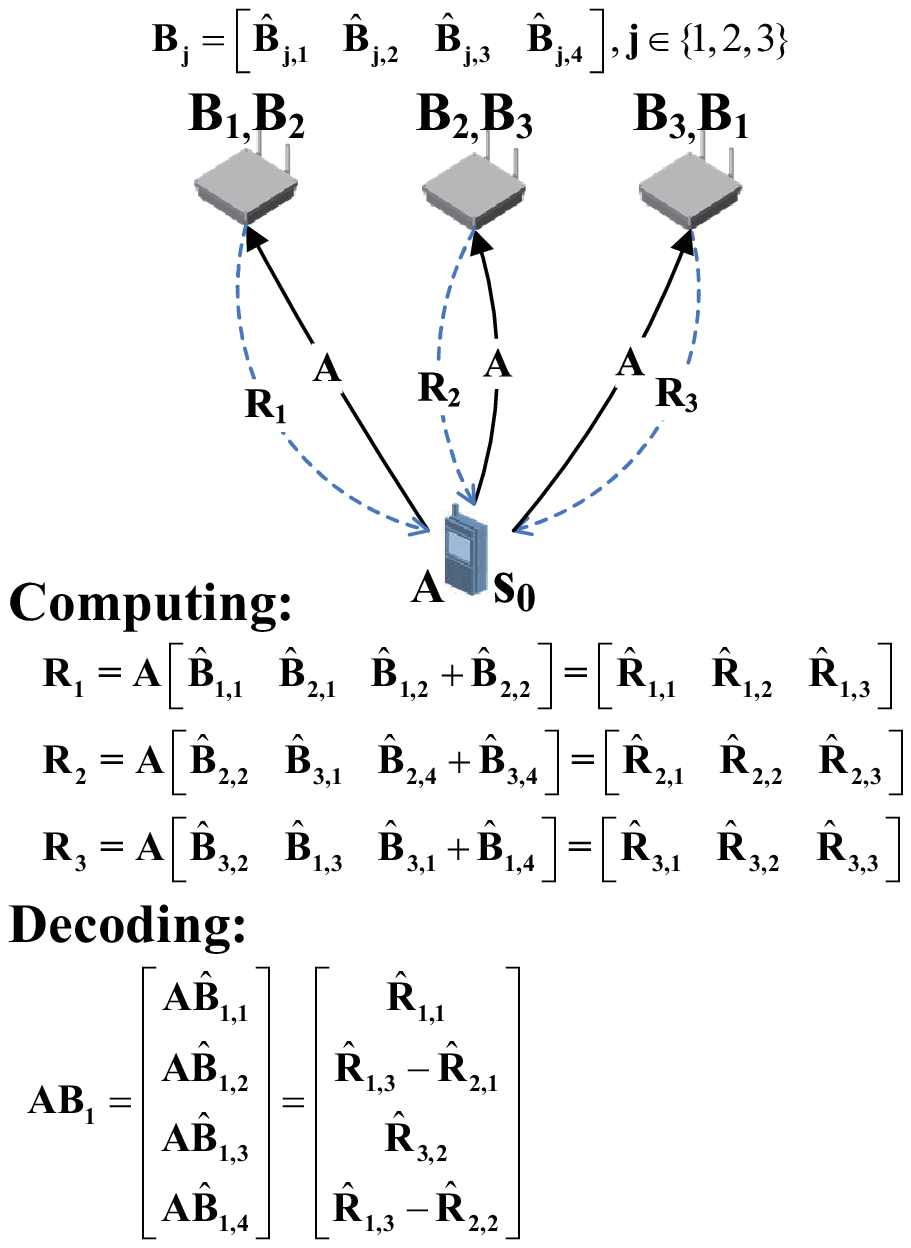}%P4.eps}

\mbox{\footnotesize (d) PCC}
\end{minipage}
\caption{Examples of distributed matrix multiplication in PEC.}
\label{Fig.2}
\end{figure*}

	Next we give examples to illustrate the impacts of storage allocation and computation design on the performance of PEC.
In \figref{Fig.2}, we have three non-colluding edge devices. The library $\textbf{B}=\{\textbf{B$_1$},\textbf{B$_2$},\textbf{B$_3$}\}$ is consisted of three matrices, each of which is called as a block and has been redundantly stored in edge devices.
The user distributes a local matrix $\textbf{A}$ to edge devices, and wants to obtain $\textbf{A}\textbf{B$_{1}$}$. The privacy requirement is that each edge device cannot identify which block is user's target. We use the data from all edge devices to user to measure the communication load and set the size of $\textbf{AB$_1$}$ as unit 1 for the ease of description.
 In \figref{Fig.2}, there are four private computing schemes:
\begin{itemize}
		\item \emph{Replication based Private Computing} (RPC) scheme where each edge device directly multiplies $\textbf{A}$ with all blocks it stores. After receiving $\textbf{R$_1$},\textbf{R$_2$},\textbf{R$_3$}$, the user can revive $\textbf{AB$_1$}$ as shown in the decoding process. The total communication load is $6$.

		\item  \emph{Vandermonde matrix based Private Computing} (VPC) scheme where each edge device divides each block into $3$ segments and uses the Vandermonde matrix to encode them before multiplying the result with $\textbf{A}$. After receiving $\textbf{R$_1$},\textbf{R$_2$},\textbf{R$_3$}$, the user can revive $\textbf{AB$_1$}$ as shown in the decoding process. The total communication load is $3$.
		
		\item GPC scheme where each edge device divides each block into $2$ segments, only uses part of segments and multiplies them with $\textbf{A}$. After receiving $\textbf{R$_1$},\textbf{R$_2$},\textbf{R$_3$}$, the user can revive $\textbf{AB$_1$}$ as shown in the decoding process. The total communication load is $3$.

		\item  PCC scheme where each edge device divides each block into $4$ segments and only encode part of the segments before multiplying the result with $\textbf{A}$. After receiving $\textbf{R$_1$},\textbf{R$_2$},\textbf{R$_3$}$, the user can revive $\textbf{AB$_1$}$ as shown in the decoding process. The total communication load is $\frac{9}{4}$.
	\end{itemize}

	In \figref{Fig.2} (a) - (d), every edge device selects same amount of data from each block it stores, to complete the computing task. Hence, edge devices cannot identify user's target in all schemes. Moveover, we have following observations. (1) Compared with RPC, GPC and VPC reduce the communication load by $50\%$, while PCC reduces that by $62.5\%$ under the same storage situation. It's because they divide the blocks into smaller segments, then only use part of segments or encode the segments in the computing process. It implies we can complete PEC with lower communication load by carefully design the computing scheme.
(2) Although GPC and VPC achieve the same communication loads in the example, GPC can
be applied in general case while VPC can only be applied in special cases. For example, there are only two edge devices, who store $\textbf{B}_1,\textbf{B}_2$ and $\textbf{B}_3,\textbf{B}_1$ respectively. Then in VPC, we will have $\textbf{R}_1=\textbf{A}[\overline{\textbf{B}}_{1,1}+\overline{\textbf{B}}_{2,1},\overline{\textbf{B}}_{1,2}+2\overline{\textbf{B}}_{2,2},\overline{\textbf{B}}_{1,3}+4\overline{\textbf{B}}_{2,3}]$, $\textbf{R}_2=\textbf{A}[\overline{\textbf{B}}_{3,1}+\overline{\textbf{B}}_{1,1},\overline{\textbf{B}}_{3,2}+2\overline{\textbf{B}}_{1,2},\overline{\textbf{B}}_{3,3}+4\overline{\textbf{B}}_{1,3}]$, where we cannot decode all segments of $\textbf{AB}_1$. We will show specific solvable conditions in \secref{EDC}. However, in GPC we will have $\textbf{R}_1=\textbf{A}[\textbf{B}_1,\textbf{B}_2],\textbf{R}_2=\textbf{A}[\textbf{B}_3,\textbf{B}_1]$ and can recover $\textbf{AB}_1$. It's worth noting that the communication load of GPC changes to $4$ with the change of storage situation. It implies that storage allocation affects the performance of computing schemes.

In this paper, we focus on the joint research of storage allocation and computation design in EC scenario, under the circumstance of (1) completing the computing tasks, (2) protecting the user's privacy, and (3) decreasing the communication load.
According to the author's knowledge, such a PEC problem has not been studied through jointly studying the storage allocation and computing schemes. The main contributions of this paper are summarized as follows:

\begin{itemize}

		\item We jointly study storage allocation and computation design in an unified framework to achieve private matrix multiplication in EC scenario, where edge devices have storage limit. To this end, we formally define the PEC problem.%fully utilize the computing resources in edge device to achieve private matrix multiplication in EC scenario, where edge devices have storage limit. To this end, we formally define the PEC problem.

		\item We fully utilize the redundant storage resources in edge devices and give a general storage allocation scheme to distribute the library to them.
		
		\item We design two computing schemes to protect the user's privacy with lower
communication load. Moreover, we conduct solid theoretical analysis on them and compare them with other private schemes in detail.

		\item We conduct extensive simulation experiments to demonstrate the effectiveness of the proposed computing schemes.
		
	\end{itemize}
	
	The rest of the paper is organized as follows. We first introduce system model in \secref{sec.sys}. Then we design an efficient scheme for PEC, including the storage allocation and private computation design in \secref{Sec.alg}. Next we give the theoretical analysis of the proposed computing schemes in \secref{TA} and compare them with other private schemes. Simulations are conducted in \secref{sec.sim}. Finally, we conclude the paper in \secref{sec.conc}.

\section{Problem Modeling}
	\label{sec.sys}

In this section, we first introduce the EC model and then present the attack model considered in this paper. Then, we give the formal definition of the PEC problem and provide an overview of the framework to solve the problem.
	
	\subsection{System Model}	\label{sec.sm}
	In this paper, we focus on the EC model which is composed of three parts: cloud, user and edge devices \cite{Cao2019,CDC2,Zhou2019,Fu2019}.
The cloud is trustworthy and has powerful computing capability.
Moreover, it keeps a library $\textbf{B}=\{\textbf{B$_{1}$}$, $\textbf{B$_{2}$}$, $\cdots$, $\textbf{B$_{w}$}\}$, which is consisted of $w$ matrices over a finite field  $\mathbb{F}_q$, $w\geq 2$. We call such a matrix as a block and $\textbf{B$_{j}$} \in \mathbb{F}_q^{r\times s}$ is the $j$-th block in $\textbf{B}$, $j\in[w]$. Note that for $a,b\in \mathbb{Z^+},a\leq b$, the notation $[a:b]$ denotes $\{a, a+1, \cdots , b\}$ and $[b]$ is the abbreviation of $[1:b]$.
We take the parts except the cloud as $S=\{s_0, s_1, \cdots, s_n\}$, where $s_0$ denotes the user and $s_i$ represents the $i$-th edge device, $i\in [n]$, $n\geq 2$.
Every edge device can store $t_0$ blocks at most, and we call $t_0$ as the storage limit, $t_0>0$.
User holds a local matrix $\textbf{A} \in \mathbb{F}_q^{m\times r}$ and only wants to use one block in $\textbf{B}$, to obtain $\textbf{AB$_{\theta}$}$, $\theta\in [w]$.
We denote $\textbf{T$_{j}$}$ as the product of $\textbf{A}$ and $\textbf{B$_{j}$}$, $j\in [w]$.

%We study the case where the number of blocks in each edge device is the same \cite{CDC2,C3,Li2016}. We denote $t$ as the storage factor, which is the number of blocks stored in each edge device, $t\in\{\mathcal{t}|\mathcal{t}\in[w],\mathcal{t}n\geq w,\mathcal{t}\leq t_0\}$. It's worth noting that if $tn<w$, we cannot arrange all blocks to edge device. Meanwhile to full use the storage resource in edge device, each block won't be restored in one edge device, followed by $t\in[w]$. Since each block may be the computing target, we try to keep the frequency of each block appearing in edge devices the same. Then we can easily get that each block appears at least $\lfloor\frac{tn}{w}\rfloor$ times, where we denote $\alpha=\lfloor\frac{tn}{w}\rfloor$ as the redundancy factor. Since $t\geq w,tn\geq w$, we have $\alpha\in[n]$.

We study the case where the number of blocks in each edge device is the same \cite{CDC2,C3,Li2016}. We denote $t$ as the storage factor, which is the number of blocks stored in each edge device, $t\in\mathbb{Z^+}$. It's worth noting that to full use the storage resource in edge devices, each block won't be restored in one edge device, followed by $t\in[w]$.
Since each block may be the computing target, we require that each block will be stored at least once and try to keep the frequency of each block appearing in edge devices the same.
Then we can easily get that each block appears at least $\lfloor\frac{tn}{w}\rfloor$ times, where we denote $\alpha=\lfloor\frac{tn}{w}\rfloor$ as the redundancy factor. %Since $t\geq w,tn\geq w$, we have $\alpha\in[n]$.
According to $S$, $\textbf{B}$ and $t_0$, the cloud selects a specific value of $t$ to arrange the blocks. Then the cloud replicates the blocks and distributes them to edge devices. We denote $\textbf{V$_i$}=\{\textbf{B$_{i_1}$}, \textbf{B$_{i_2}$}, \cdots, \textbf{B$_{i_{t}}$}\}$ as the set of blocks in $s_i$, where $\textbf{B$_{i_u}$}$ is the $u$-th block, $u\in [t]$. Meanwhile we denote $\widetilde{V}_i$ as the set of the indexes of all $t$ blocks in $s_i$, where $\widetilde{V}_{i,j}$ is the $j$-th index.

		\begin{figure}[tb]
		\centering
		\includegraphics[width=2.6in]{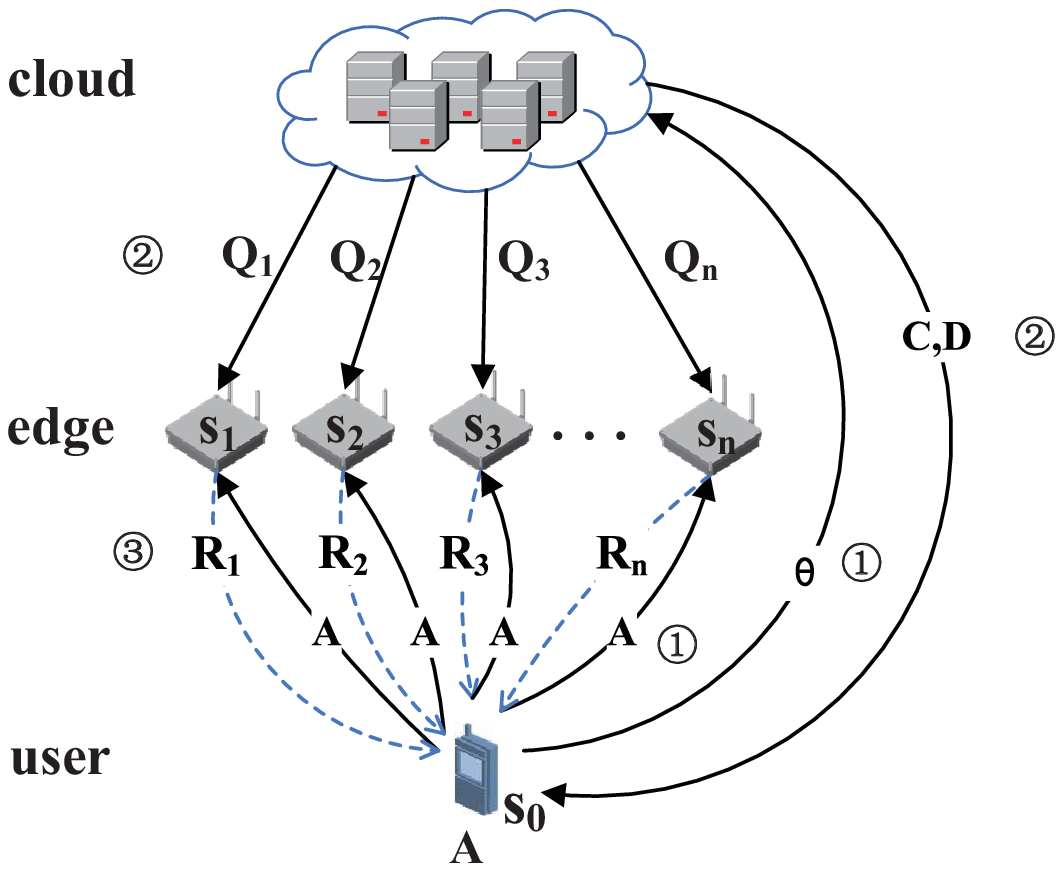}
		\caption{The overall process of private edge computing.}
		\label{Fig.3}
	\end{figure}

After the storage allocation is fixed, we can start the computing tasks.
Firstly, we denote $(\cdot)_{[i,j]}$ as the element in the $i$-th row and $j$-th column of a matrix and have the following definition:
	\begin{DD}\label{Def.BMM}(\textbf{Block Matrix Multiplication $\circledast$})
	Assuming that there is an $e\times f$ dimensional matrix $\textbf{Q}$ and a $g\times h$ dimensional matrix $\textbf{F}$. $\textbf{Q$_u$}$ is the $u$-th row of $\textbf{Q}$. We divide\footnote{If it is indivisible, we will add $\lceil\frac{h}{f}\rceil*f-h$ columns of 0 elements to the right end of the matrix.} $\textbf{F}$ into $f$ segments by column and $\textbf{F$_v$}$ is $v$-th segment.
Then the operator $\circledast$ is defined as follows:
	\begin{equation} \label{Eq.bmm}
	\begin{aligned}
\textbf{G}_u &=\textbf{Q}_u \circledast \textbf{F}=\sum\limits_{v=1}^{f} \textbf{Q}_{[u,v]}\textbf{F$_v$}, u\in[e],\\
     & \textbf{G}=\textbf{Q} \circledast\textbf{F}=[\textbf{G$_1$},\cdots,\textbf{G$_e$}],
     	\end{aligned}
		\end{equation}		
where we call $\textbf{G}_u,u\in[e]$ as an intermediate value.
	\end{DD}

\begin{figure}[htb]
		\centering
		\includegraphics[width=3in]{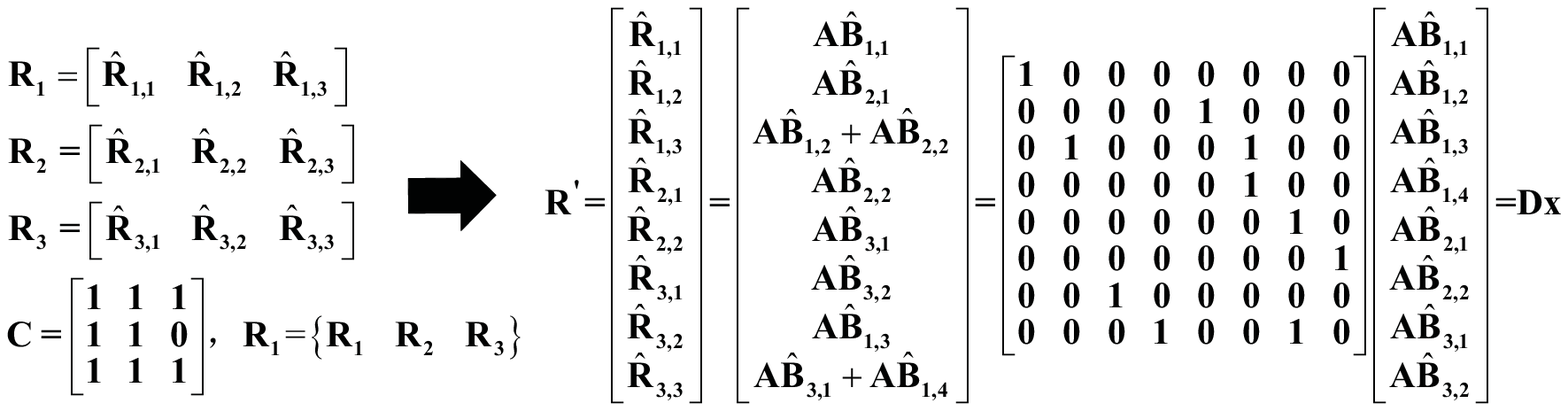}
		\caption{The specific decoding process of \figref{Fig.2} (d).}
		\label{Fig.pccdec}
	\end{figure}

Secondly, we detail the overall computing process in \figref{Fig.3}:
\begin{itemize}
		\item User $s_0$ first sends the target block's index $\theta$ to the cloud and distributes \textbf{A} to edge devices.
		
		\item After requiring $\theta$ from $s_0$, the cloud can generate the computing scheme, including a set of encoding coefficient matrices $\textbf{Q}=\{\textbf{Q$_{1}$}, \cdots, \textbf{Q$_{n}$}\}$, where $\textbf{Q$_{i}$} \in \mathbb{F}_2^{g\times lt}$ and $g,l$ are variables differ in different computing schemes, $i\in[n]$. Meanwhile for the decoding process in $s_0$, the cloud generates a selection matrix $\textbf{C}\in \mathbb{F}_2^{n\times g}$ and a decoding coefficient matrix $\textbf{D}\in\mathbb{F}_2^{h\times h}$, where $h$ is a variable differs in different computing schemes. It's worth noting that we have $\sum\limits_{i=1}^n \sum\limits_{j=1}^g \textbf{C}_{[i,j]}=h$ in our computing schemes.
Then the could sends $\textbf{C},\textbf{D}$ to $s_0$ and $\textbf{Q$_{i}$}$ to $s_i$, $i\in[n]$.

		\item After receiving $\textbf{Q$_{i}$}$ and $\textbf{A}$, edge device $s_i$ can start the computing task, $i\in[n]$.
Firstly, $s_i$ divides\footnote{If it is indivisible, we will add $\lceil\frac{s}{l}\rceil*l-s$ columns of 0 elements to the right end of the block.} each block in $\textbf{V$_i$}$ into $l$ segments by column and concatenates these segments to get $\textbf{F$_i$}= [\textbf{B$_{i_1,1}$},\cdots,\textbf{B$_{i_1,l}$},\textbf{B$_{i_2,1}$},\cdots,\textbf{B$_{i_2,l}$} \cdots, \textbf{B$_{i_t,l}$}]$, where $\textbf{B}_{i_j,v}$ is the $v$-th segment of $\textbf{B}_{i_j}$, $v\in[l],j\in[t]$.
Secondly, $s_i$ computes $\textbf{G$_{i}$}=\textbf{Q$_{i}$}\circledast \textbf{F$_i$}$, where $\textbf{G}_{i,u}$ is the $u$-th intermediate value, $u\in[g]$.
Thirdly, $s_i$ computes $\textbf{R$_{i}$}=\textbf{A}\times\textbf{G$_i$}$ and returns it to $s_0$.
We denote $\textbf{R}_{i,u}=\textbf{A}\textbf{G}_{i,u}$ as the $u$-th value in $\textbf{R$_{i}$}$.
We denote $\textbf{R}$ as the set of these answers, {\em i.e.}, $\textbf{R}=\{\textbf{R$_{1}$}, \cdots, \textbf{R$_{n}$}\}$.

	\item When receiving $\textbf{C},\textbf{D}$ and $\textbf{R}$, $s_0$ can start the decoding task.
Firstly, $s_0$ selects the useful values in $\textbf{R}$ according to $\textbf{C}\in \mathbb{F}_2^{n\times g}$. Specifically, for $\forall i\in[n]$, $s_0$ computes $b_i=\sum\limits_{j=1}^g \textbf{C}_{[i,j]}$ and $\textbf{R}'_{i}=[{\textbf{R}_{i,z_1}}^{\top},\cdots,{\textbf{R}_{i,z_{b_i}}}^{\top}]^{\top}$, where $\textbf{C}_{[i,z_k]}=1,k\in[b_i]$.
Secondly, $s_0$ can concatenate ${\textbf{R}'_{i}}$ to get $\textbf{R}'=[{\textbf{R}'_{1}}^{\top},\cdots,{\textbf{R}'_{n}}^{\top}]^{\top}$.
We use the example in \figref{Fig.2} (d) to detail the decoding process and show the detail in \figref{Fig.pccdec}. After receiving $\textbf{C}$ and $\textbf{R}$, $s_0$ calculate $\textbf{R}'$ as shown in \figref{Fig.pccdec}. We find that there are $h=\sum\limits_{i=1}^n \sum\limits_{j=1}^g \textbf{C}_{[i,j]}$ values in $\textbf{R}'$, each of which is a linear combination of $h$ specific segments from $\{\textbf{T}_{j,v}|j\in[w],v\in[l]\}$. Meanwhile $\textbf{D}$ is the encoding coefficient matrix of them. We take a $1\times h$ dimensional vector $\textbf{x}$ to represent these segments, followed by $\textbf{R}'=\textbf{D}\textbf{x}$. %It is worth noting that for the recovery of $\textbf{T$_{\theta}$}$, all segments of it will appear in $\textbf{x}$.
Thirdly, $s_0$ can decode $\textbf{x}$ from $\textbf{R}'=\textbf{D}\textbf{x}$ to recover $\textbf{T$_{\theta}$}$.
	\end{itemize}

 	%In our study, without loss of generality, we focus on the multiplication of two matrices. The schemes proposed in this paper can also be applied to other cases that require multiplication of data matrix with one input vector and/or multiplication of a data matrix with different input vectors.
	
	To ensure that user can recover $\textbf{T$_{\theta}$}$, we give the availability condition:
	\begin{DD}\label{Def.Availability}(\textbf{Availability Condition})
	The user can decode $\textbf{R}'=\textbf{D}\textbf{x}$ to recover $\textbf{T$_{\theta}$}$ iff $Rank(\textbf{D})=h$.
	\end{DD}

	In this paper, we assume that all edge devices are available, {\em i.e.}, all results will be correctly computed and transmitted to the user in a timely manner. Meanwhile we consider the consumption of storage and communication resources of proposed computing schemes. Firstly, we define the storage load $s_d$ as the number of elements in $\textbf{V$_{i}$}$, normalized by the size of $\textbf{B$_\theta$}$. Hence $s_d=t$. Secondly, the communication traffic between edge devices and user is consisted of the upload and download traffic. Because the upload traffic from user to every edge device is fixed as the size of $\textbf{A}$, so the communication traffic is mainly determined by the download traffic. Hence, we only consider the download traffic.
We define the communication load $c_i$ as the number of elements in $\textbf{R$_{i}$}$, normalized by the size of $\textbf{T$_\theta$}$. Hence, the communication load of all edge devices is: $c_{d}=\sum\limits_{i=1}^n c_i$.
\iffalse
	{\chr
		{	\begin{eqnarray}
			\begin{aligned}
			&c_{d}=\sum\limits_{i=1}^n c_i.
			\end{aligned}
			\label{cost}
			\end{eqnarray}
		}
}
\fi

\begin{table}
		\caption{Notations}
		\center
		\begin{tabular}{|c|p{7cm}|}
			\hline \begin{minipage}{1cm}\vspace{1mm} \textbf{Notations}\vspace{1mm} \end{minipage} & \textbf{Meaning }\\
			\hline
			\begin{minipage}{1cm}\vspace{1mm} $\textbf{A}$\vspace{1mm} \end{minipage}& \parbox {2.7in}{the local data matrix and $\textbf{A}\in \mathbb{F}_q^{m\times r}$.}\\
			\hline \begin{minipage}{1cm}\vspace{1mm} $[a:b]$\vspace{1mm} \end{minipage}& \parbox {2.7in}{it denotes $\{a, a+1, \cdots , b\}$ and $[b]$ is the abbreviation of $[1:b]$, where $a,b\in \mathbb{Z^+},a\leq b$.}\\
			\hline
			\begin{minipage}{1cm}\vspace{1mm} $\alpha$\vspace{1mm} \end{minipage}& \parbox {2.7in}{ the redundancy factor, where each block appears in at least $\alpha$ edge devices, $\alpha=\lfloor\frac{tn}{w}\rfloor$.}\\
			\hline
			\begin{minipage}{1cm}\vspace{1mm} $\textbf{B}$\vspace{1mm} \end{minipage}& \parbox {2.7in}{the set of $w$ data matrices $\textbf{B}=\{\textbf{B$_{1}$}, \textbf{B$_{2}$}, \cdots, \textbf{B$_{w}$}\},w\geq2$, $\textbf{B$_{j}$}\in \mathbb{F}_q^{r\times s}$ and will be further divided to $l$ segments by column.}\\
			\hline
			\begin{minipage}{1cm}\vspace{1mm} $\textbf{B$_{j,v}$}$\vspace{1mm} \end{minipage}& \parbox {2.7in}{the $v$-th segments in $\textbf{B$_{j}$}$, $v \in[l]$.}\\
			%\hline
			%\begin{minipage}{1cm}\vspace{1mm} $\textbf{B$_{i_u}$}$\vspace{1mm} \end{minipage}& \parbox {2.7in}{the $u$-th blocks in $s_i$, $i \in [n], u\in [t]$.}\\
			\hline
			\begin{minipage}{1cm}\vspace{1mm} $\textbf{B}'_{j}$\vspace{1mm} \end{minipage}& \parbox {2.7in}{the rearranged block, which stems from $\textbf{B$_{j}$}$ and $\textbf{P}$,\newline $\textbf{B}'_{j}=[\textbf{B}_{j,\textbf{P}_{[j,1]}},\cdots,\textbf{B}_{j,\textbf{P}_{[j,l]}}]$.}\\
			%\hline
			%\begin{minipage}{1cm}\vspace{1mm} $\beta$\vspace{1mm} \end{minipage}& \parbox {2.7in}{for given $\alpha\in[2,n]$, $\beta=\{\mathcal{t}|\lfloor\frac{\mathcal{t}n}{w}\rfloor=\alpha,\mathcal{t}\in\mathbb{Z^+}\}$.}\\
			\hline
			\begin{minipage}{1cm}\vspace{1mm} $\textbf{C}$\vspace{1mm} \end{minipage}& \parbox {2.7in}{the selection matrix used in $s_0$, $\textbf{C}\in \mathbb{F}_2^{n\times g}$.}\\
			\hline
			\begin{minipage}{1cm}\vspace{1mm} $\textbf{D}$\vspace{1mm} \end{minipage}& \parbox {2.7in}{the decoding coefficient matrix used in $s_0$, $\textbf{D}\in\mathbb{F}_2^{h\times h}$.}\\
			\hline
			\begin{minipage}{1cm}\vspace{1mm} $\textbf{F}_i$\vspace{1mm} \end{minipage}& \parbox {2.7in}{the concatenation of all segments in $s_i$, \newline $\textbf{F$_i$}=[\textbf{B$_{i_1,1}$},\cdots,\textbf{B$_{i_1,l}$},\textbf{B$_{i_2,1}$},\cdots,\textbf{B$_{i_2,l}$} \cdots, \textbf{B}_{i_t,l}]$.}\\
			\hline
			\begin{minipage}{1cm}\vspace{1mm} $\textbf{M}$\vspace{1mm} \end{minipage}& \parbox {2.7in}{A general expression for the $t$ product blocks stored by each edge device, $\textbf{M}=\{\textbf{M$_{1}$}, \textbf{M$_{2}$}, \cdots, \textbf{M$_{t}$}\}$.}\\
			\hline
			\begin{minipage}{1cm}\vspace{1mm} $\textbf{G}_{i}$\vspace{1mm} \end{minipage}& \parbox {2.7in}{$\textbf{G$_{i}$}=\textbf{Q$_{i}$}\circledast \textbf{F$_i$},i\in[n]$.}\\
			\hline
			\begin{minipage}{1cm}\vspace{1mm} $p$\vspace{1mm} \end{minipage}& \parbox {2.7in}{$p=\alpha^{t-2}+(\alpha-1)^{t-1}$.}\\
			\hline
			\begin{minipage}{1cm}\vspace{1mm} $\textbf{$\textbf{P}$}$\vspace{1mm} \end{minipage}& \parbox {2.7in}{ a $w\times l$ dimensional random matrix, which is used to select a segment from product blocks randomly.}\\
			\hline
			\begin{minipage}{1cm}\vspace{1mm} $\textbf{Q}$\vspace{1mm} \end{minipage}& \parbox {2.7in}{the set of encoding coefficient matrices,\newline $\textbf{Q}=\{\textbf{Q$_{1}$}, \textbf{Q$_{2}$}, \cdots, \textbf{Q$_{n}$}\},\textbf{Q$_{i}$} \in \mathbb{F}_2^{g\times lt},i\in[n]$.}\\
			\hline
			\begin{minipage}{1cm}\vspace{1mm} $\textbf{R}$\vspace{1mm} \end{minipage}& \parbox {2.7in}{the set of answer matrices the edge devices send to the user,\newline $\textbf{R}=\{\textbf{R$_{1}$}, \textbf{R$_{2}$}, \cdots, \textbf{R$_{n}$}\}, \textbf{R$_{i}$}=\textbf{A}\times \textbf{G$_i$},i\in[n]$.}\\
			\hline
			\begin{minipage}{1cm}\vspace{1mm} $\textbf{R}'$\vspace{1mm} \end{minipage}& \parbox {2.7in}{the concatenation of $\textbf{R$_{i}$}'$, which stems from $\textbf{R$_{i}$}$ and $\textbf{C}$,\newline $\textbf{R}'=[{\textbf{R}'_{1}}^{\top},\cdots,{\textbf{R}'_{n}}^{\top}]^{\top}$.}\\
			\hline
			\begin{minipage}{1cm}\vspace{1mm}
				$Rank(\cdot)$ \end{minipage}& \parbox {2.7in}{the rank of a vector set or matrix.}\\
			\hline
			\begin{minipage}{1cm}\vspace{1mm} $S$\vspace{1mm} \end{minipage}& \parbox {2.7in}{the set of a user and $n$ edge devices, $S=\{s_0, s_1\cdots, s_n\},n\geq2$.}\\
			\hline
			\begin{minipage}{1cm}\vspace{1mm} $\textbf{T$_{j}$}$ \vspace{1mm} \end{minipage}& \parbox {2.7in}{the product of local data $\textbf{A}$ and data matrix $\textbf{B$_{j}$},j \in[w]$}.\\
			\hline
			\begin{minipage}{1cm}\vspace{1mm} $t$\vspace{1mm} \end{minipage}& \parbox {2.7in}{ the storage factor, where each edge device stores $t$ blocks, $t\in[2:w],tn\geq w,t\leq t_0$.}\\
			\hline
			\begin{minipage}{1cm}\vspace{1mm} $t_0$\vspace{1mm} \end{minipage}& \parbox {2.7in}{ the storage limit, where each edge device can store $t_0$ blocks at most, $t_0\geq2$.}\\
			\hline
			\begin{minipage}{1cm}\vspace{1mm} $\textbf{V$_i$}$\vspace{1mm} \end{minipage}& \parbox {2.7in}{the set of blocks stored in $s_i$, $\textbf{V$_i$}=\{\textbf{B$_{i_1}$}, \textbf{B$_{i_2}$}, \cdots, \textbf{B$_{i_{t}}$}\}$ where \textbf{B$_{i_u}$} is the $u$-th blocks, $i\in [n],u\in[t]$}\\
			\hline
			\begin{minipage}{1cm}\vspace{1mm} $\widetilde{V}_i$\vspace{1mm} \end{minipage}& \parbox {2.7in}{the set of the indexes of the $t$ blocks in $s_i,i \in [n]$, where $\widetilde{V}_{i,j}$ is the $j$-th index.}\\

			\hline
			\begin{minipage}{1cm}\vspace{1mm} $(\cdot)^{\top}$\vspace{1mm} \end{minipage}& \parbox {2.7in}{the transposition of matrix.}\\
			\hline
			\begin{minipage}{1cm}\vspace{1mm} $(\cdot)_{[i,j]}$\vspace{1mm} \end{minipage}& \parbox {2.7in}{the element in the $i$-th row and $j$-th column of a matrix.}\\

			%			\hline		
			%			 \begin{minipage}{1cm}\vspace{1mm}
			%			 $\textbf{E}_{t}$ \vspace{1mm} \end{minipage}& \parbox {2.7in}{the $t\times t$ dimensional identify matrix.}\\
			%			\hline
			%             \begin{minipage}{1cm}\vspace{1mm}
			%			 $\textbf{O}_{p,q}$ \vspace{1mm} \end{minipage}& \parbox {2.7in}{the $p\times q$ dimensional zero matrix.}\\
			\hline	
		\end{tabular}\label{tab:notation}
	\end{table}

	To facilitate the discussions, we define notations in \tabref{tab:notation}.

	\subsection{Attack Model and Privacy Requirements}\label{Sec.Attack}
	In this paper, we study the passive attack model where every edge device may be a passive attacker or compromised by a passive attacker. Moreover, they don't collude with each other. A similar passive attack model has been investigated in secure distributed computing \cite{C9,C14,C15,Cao2019,M3,M9}.
	Edge devices may want to know the index of the target block, which compromises user's privacy. It worth noting that if $t=1$, each edge device stores one block and there is no solution for the privacy protection. Hence, we discuss $t\geq2 ,t_0\geq2$ in following paper.

	User's privacy is protected when none of the edge devices can identify $\theta$, after the user recovers $\textbf{T}_\theta$. The privacy condition \cite{M1,M5,M9} is defined as follows:
	\begin{DD} \label{Def.Privacy}(\textbf{Privacy Condition})
		A computing scheme satisfies the privacy requirements iff
		\begin{equation} \label{Eq.Privacy}
		\textbf{I}(\theta;\textbf{Q$_{i}$},\textbf{A},\textbf{R}_i,\textbf{V$_i$})=0, \forall{i}\in [n],
		\end{equation}		
in which $\textbf{I}(\cdot ; \cdot)$ is the mutual information.
	\end{DD}

We denote $\mathcal{T}=\{\mathcal{t}|\mathcal{t}\in[2:w],\mathcal{t}n\geq w,\mathcal{t}\leq t_0\}$, then we have the following lemma:

\begin{LL}\label{Lem_t}
		If there exists a storage allocation scheme for PEC problem, we have $t\in\mathcal{T}$.
	\end{LL}
	\begin{proof}
Firstly, we have $t\in[w]$ as defined in \secref{sec.sm}. Secondly, as mentioned above, we discuss $t\geq2 ,t_0\geq2$ in following paper. Thirdly, since each edge device can store $t_0$ blocks at most as mentioned in \secref{sec.sm}, we have $t\leq t_0$.
Meanwhile, from the \secref{sec.sm}, we find that there are $n$ edge devices and each of them stores $t$ blocks. Since we  require that each block will be stored at least once, we have $tn\geq w$.
Therefore, in our storage allocation scheme of PEC, we have $t\in\mathcal{T}$.
	\end{proof}
The private computing scheme is based on the storage allocation scheme. Hence, in PEC we only discuss $t\in\mathcal{T}$ in following paper.

	\subsection{Problem Definition}

	The \emph{Private Edge Computing} (PEC) problem is defined as follows:
	\begin{DD} \label{Def.FRTPCEC}
		Given an EC system $S$, a library $\textbf{B}$ and the storage limit $t_0$, the PEC problem is to distribute the blocks in $\textbf{B}$ to edge devices, i.e., storage allocation and design a computing scheme, i.e., computation design, that satisfies the privacy and availability conditions.
	\end{DD}

	\subsection{The PEC Framework} \label{Sec.framework}
	In this section, we provide an overview of the framework to solve the PEC problem where the key components in the framework will be elaborated in the following sections.
	\begin{itemize}
		\item \textbf{Storage Allocation.} In this step, the cloud should give a general storage allocation scheme to distribute the library to edge devices for any possible value of the storage factor $t$. We will present the general storage allocation scheme in \secref{SA}.
			
		\item \textbf{Private Computation Design.}
		The design of the computing scheme, which satisfies the privacy and availability conditions, is mainly on the cloud. The cloud shall generate a set of encoding coefficient matrices $\textbf{Q}$ for edge devices. Meanwhile the cloud generates a selection matrix $\textbf{C}$ and a decoding coefficient matrix $\textbf{D}$ for $s_0$.
We will elaborate the design of $\textbf{Q}$, $\textbf{C}$ and $\textbf{D}$ in \secref{PLC}.

		\item \textbf{Coded Edge Computing.}
	    After receiving $\textbf{Q$_{i}$}$ from the cloud and $\textbf{A}$ from $s_0$, $s_i$ can start the computing tasks. $s_i$ first divides each block it stores into $l$ segments by column and concatenates these segments to get $\textbf{F$_i$}$. Then $s_i$ computes $\textbf{G$_{i}$}=\textbf{Q$_{i}$}\circledast \textbf{F$_i$}$. Finally, $s_i$ computes $\textbf{R$_{i}$}=\textbf{A}\times\textbf{G$_i$}$ and returns it to $s_0$. It's worth noting that since $\textbf{F$_i$}$ is the concatenation of $lt$ segments in $s_i$, $\textbf{G$_{i}$}$ is actually a linear combination of these segments, which implies the design of $\textbf{Q$_{i}$}$ is equal to arranging the segments in $s_i$ to get $\textbf{G$_{i}$}$.

		\item \textbf{Target Result Recovery.}
When receiving $\textbf{C},\textbf{D}$ and $\textbf{R}$, $s_0$ can start the decoding task. Firstly, $s_0$ selects the useful values in $\textbf{R$_{i}$}$ according to $\textbf{C}$ to get $\textbf{R$_{i}$}',i\in[n]$. Secondly, $s_0$ can concatenate $\textbf{R}'_{i}$ to get $\textbf{R}'$. Thirdly, $s_0$ can decode $\textbf{x}$ from $\textbf{R}=\textbf{D}\textbf{x}'$ to recover $\textbf{T$_{\theta}$}$.
	\end{itemize}

\section{The PEC Schemes}
	\label{Sec.alg}
	In this section, we present a general storage allocation scheme for any possible value of the storage factor $t$. Then based on the storage allocation, we give two private computing schemes respectively.

	\subsection{Storage Allocation}\label{SA}
	%Since the selection of $tt\in\{\mathcal{t}|\mathcal{t}\in[2:w],\mathcal{t}n\geq w,\mathcal{t}\leq t_0\}$ is related with the fundamental resources consumption of the coding schemes, we will elaborate it in the following sections.
For given storage factor $t$, there are two problems to be solved in the storage allocation: (1) each block will be replicated how many times, (2) each block will be distributed to which edge device. Next we give the general storage allocation scheme from above two parts.
	As mentioned in \secref{SA}, there are $w$ blocks in $\textbf{B}$ and we try to keep the frequency of each block appearing in edge devices the same. When $t$ is fixed, we have that there are $n$ edge devices and each of them stores $t$ blocks. Hence, we can easily get that there are $p_r=tn-w\alpha$ blocks appear $\lceil\frac{tn}{w}\rceil$ times and the other blocks appear $\alpha$ times, $\alpha=\lfloor\frac{tn}{w}\rfloor$. Without loss of generality, we replicate the first $p_r$ blocks $\lceil\frac{tn}{w}\rceil$ times and the other blocks $\alpha$ times. Since $\alpha\leq \lceil\frac{tn}{w}\rceil\leq \alpha+1$, each block appears in at most $\alpha+1$ edge devices.

	Next we can arrange the above $tn$ blocks to $n$ edge devices. We denote $\textbf{O}$ as the set of these $tn$ blocks, where the blocks are sorted in ascending order of their indexes and $\textbf{O}_j$ is the $j$-th block in $\textbf{O}$, $j\in[tn]$. Then for $\textbf{O}_j$, if $j$ is divisible by $n$, we will distribute it to $s_n$. Otherwise, we will distribute it to the $j\%n$-th edge device. Therefore, we have $\textbf{V$_i$}=\{\textbf{O}_i,\textbf{O}_{n+i},\cdots,\textbf{O}_{n(t-1)+i}\}$.

\begin{figure}[htbp]
		\centering
		\includegraphics[width=1.9in]{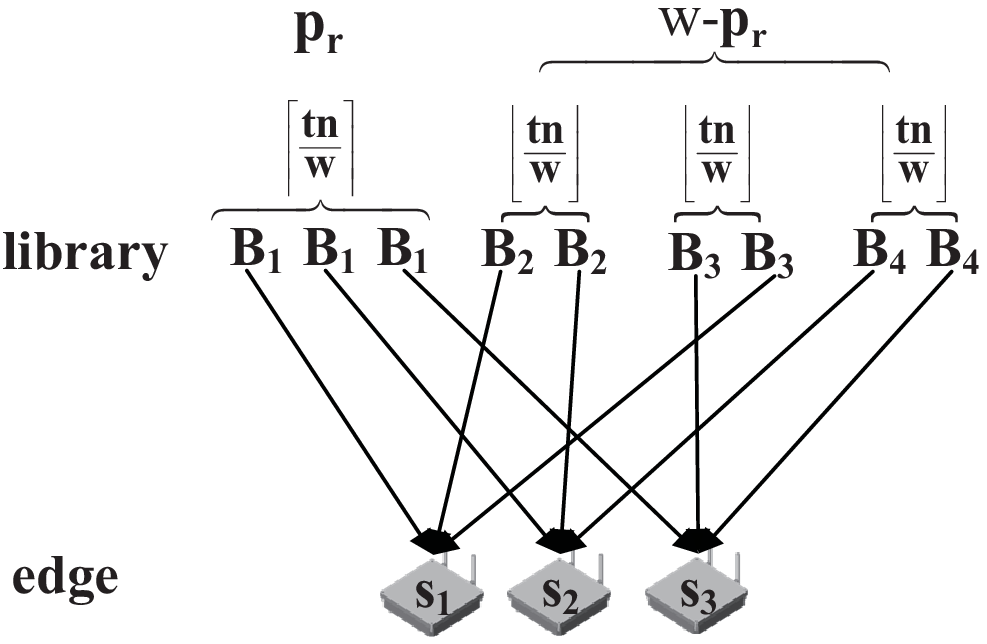}
		\caption{An example of the storage allocation scheme.}
		\label{Fig.5}
	\end{figure}

	\figref{Fig.5} shows an example of the storage allocation scheme where $n=3,w=4,t=3$. The first $p_r=1$ block appears $\lceil\frac{tn}{w}\rceil=3$ times and the others appear $\alpha=2$ times. Then we sort them in ascending order of their indexes and arrange the blocks to edge devices in order.

	\subsection{Private Computation Design}\label{PLC}
	After storage allocation phase, $s_i$ stores $t$ blocks and $\textbf{B}_j$ appears in at least $\alpha$ edge devices, $i\in[n],j\in[w]$. Then in this subsection, we give the design of two private computing schemes. The first is the \emph{General Private Computation} (GPC) scheme which can be applied in general case. The second is the \emph{Private Coded Computation} (PCC) scheme, which can only be applied in special cases but achieves less communication load.

	We note that the design of the computing scheme including two parts: (1) the design of the encoding coefficient matrix $\textbf{Q$_{i}$}\in\mathbb{F}_2^{g\times lt}$, which is equal to arranging all segments in $s_i$ to get $\textbf{G$_i$}$, (2) the design of $\textbf{C}\in\mathbb{F}_2^{n\times g},\textbf{D}\in\mathbb{F}_2^{h\times h}$, which is equal to determining the specific segments to decode and selecting values used in the decoding process. Next we present the two schemes from above two aspects.

	\subsubsection{GPC}\label{sGPC}
	 Firstly, we give the design of encoding process. It's easy for us to come up with an idea to distribute the task of computing $\textbf{T$_\theta$}$ by asking the edge devices, that store $\textbf{B$_\theta$}$, to compute part of it. %Meanwhile, except use $\textbf{B$_\theta$}$, we have to use other blocks for the privacy requirement.
Specifically, we first let $l=\alpha$. Then for $s_i$, we directly select one segment from each block it stores as an intermediate value. Hence, there are $t$ intermediate values in $\textbf{G$_i$}$. Since  $\textbf{G$_i$}=\textbf{Q$_{i}$}\circledast\textbf{F$_i$}$, there are $t$ rows in $\textbf{Q$_{i}$}$ from \ddref{Def.BMM}, followed by $g=t$.
There are two points worth noting.
First, we have to select the segments randomly for the privacy requirement. Hence, we generate a $w\times l$ dimensional random matrix $\textbf{$\textbf{P}$}$, where the $j$-th row $\textbf{P}_j$ can rearrange the segments of $\textbf{B$_j$}$ to get the rearranged block $\textbf{B}'_j$. Then for the $v$-th time we select a segment from $\textbf{B}_j$, we will use $\textbf{B}'_{j,v}$, {\em i.e.}, $\textbf{B}_{j,\textbf{P}_{[j,v]}}$.
For example $\textbf{B}_1=[\textbf{B}_{1,1},\textbf{B}_{1,2},\textbf{B}_{1,3}]$ and $\textbf{P}_1=[2,3,1]$, we have $\textbf{B}'_1=[\textbf{B}'_{1,1},\textbf{B}'_{1,2},\textbf{B}'_{1,3}]=[\textbf{B}_{1,2},\textbf{B}_{1,3},\textbf{B}_{1,1}]$. If we want to select a segment from $\textbf{B}_1$ for the first time, we will use $\textbf{B}'_{1,1}$, {\em i.e.}, $\textbf{B}_{1,2}$.
Second, for each block $\textbf{B$_j$}$, we will select a segment of it in $s_i,i\in \{\mathcal{i}|j\in \widetilde{V}_\mathcal{i}\}$. Since there are at least $\alpha$ edge devices store $\textbf{B}_j$, we select at least $\alpha$ segments from $\textbf{B}_j$. If $j=\theta$, we require the $\alpha$ segments are totally different for the recovery of $\textbf{T$_\theta$}$.
Otherwise, the $\alpha$ segments could be the same or different with each other. Hence, without loss of generality, we set that they are the same. In other word, we always use one specific segment of $\textbf{B}_j, j\not =\theta$.

Secondly, we give the design of decoding process.
Since each intermediate value is a segment alone, we can directly select $\alpha$ different intermediate values related with $\textbf{B$_\theta$}$ for $s_0$.
Specifically, for $s_i,i\in \{\mathcal{i}|\theta\in \widetilde{V}_\mathcal{i}\}$, we select the intermediate value related with $\textbf{B$_\theta$}$. For the other edge devices, we won't select any intermediate value.
It's worth noting that if there are $\alpha+1$ edge devices that store $\textbf{B$_\theta$}$, then for the last edge device of them, we won't select any intermediate value from it.
At last, there are $\alpha$ selected intermediate values and each of them is consisted of $\alpha$ segments in $\textbf{B$_\theta$}$. Hence, $\textbf{x}=[\textbf{T$_{\theta,1}$}^\top,\cdots,\textbf{T$_{\theta,\alpha}$}^\top]^\top$.
Meanwhile, as mentioned in \secref{sec.sm}, there are $h$ values in $\textbf{R}'$, each of which is a linear combination of $h$ specific segments.
Therefore, we have $h=\alpha$, {\em i.e.}, $\textbf{D}\in\mathbb{F}_2^{\alpha \times \alpha}$.
Finally, we have that $\textbf{Q$_{i}$} \in \mathbb{F}_2^{t\times lt}$, $\textbf{C}\in \mathbb{F}_2^{n\times t}$ and $\textbf{D}\in\mathbb{F}_2^{\alpha \times \alpha}$ in GPC.

Thirdly, based on the design of above two processes, we can give the algorithm of GPC. In \algref{gPC}, $f$ is a variable to record the number of used segments in $\textbf{B}_\theta$ and $loc()$ is a subfunction to locate $\textbf{B$_j$}$ in $\textbf{V$_i$}$, followed by $loc(i,j)=b$, where $\widetilde{V}_{i,b}=j,b\in[t]$.
We first initialize the parameters (line 1-2). Then we design $\textbf{Q$_i$}$ for edge devices in turn (line 3-8). We first check the index of each block in $s_i$. If it's $\theta$ and $f<\alpha+1$, we use the $f$-th segment of $\textbf{B}'_\theta$ and update $\textbf{Q$_i$}$, $\textbf{C}$, $\textbf{D}$ (line 5 to 6). Otherwise, we use the first segment of $\textbf{B}'_j$ and update $\textbf{Q$_i$}$ (line 7 to 8). Finally we can get the whole $\textbf{Q}$, $\textbf{C}$ and $\textbf{D}$.
Lines 3 to 8 of \algref{gPC} are looped at most $nt$ times and the complexity of line 8 is $O(t)$. The complexity of line 9 is $O(n)$. Therefore, the complexity of \algref{gPC} is $O(nt^2)$.

\setlength{\textfloatsep}{0.1cm}
	\setlength{\floatsep}{0.1cm}
	\begin{algorithm}[t]%\label{alg2}
		\LinesNumbered
		\begin{small}
			\KwIn{$\theta,S,\textbf{B},\{\widetilde{V}_i|i\in[n]\}$}%
			\KwOut{$\textbf{Q}$, $\textbf{C}$, $\textbf{D}$}
			Initialize $\textbf{Q$_i$}$, $\textbf{C}$, $\textbf{D}$ as zero matrices\;
			$f=1,\alpha=\lfloor\frac{tn}{w}\rfloor$\;
			\For{$i=1$ to $n$}{
			        \For{$j\in \widetilde{V}_i$}{
 			          \uIf{$j=\theta$ and $f<\alpha+1$}{
   			               %$f=f\%\alpha+1$\;
 			               %$Q_i\gets Q_i\cup (j,f)$\;
 			               $\textbf{Q$_i$}_{[loc(i,j),(loc(i,j)-1)l+\textbf{P}_{[j,f]}]}=1$,
  			               $\textbf{C}_{[i,loc(i,j)]}=1$,
   			               $\textbf{D}_{[f,\textbf{P}_{[j,f]}]}=1$,
   			               $f=f+1$\;
 			          }
   			          \Else{
 			               $\textbf{Q$_i$}_{[loc(i,j),(loc(i,j)-1)l+\textbf{P}_{[j,1]}]}=1$\;
   			          }
			        }
			  }
			
 			 $\textbf{Q}\gets \bigcup_{i=1}^{n}\textbf{Q$_i$}$\;
			\Return $\textbf{Q}$, $\textbf{C}$, $\textbf{D}$.
			\caption{GPC Algorithm } \label{gPC}
		\end{small}
	\end{algorithm}

Next we will present the design of PCC, which may further reduce the communication traffic by combining the segments when $\alpha\geq 2$.

	\subsubsection{PCC}\label{IPC}

	Firstly, we give the design of encoding process. Since $\textbf{G$_i$}$ is actually a combination of all segments in $s_i$ as mentioned in \secref{Sec.framework}, the design of computing scheme in $s_i$ is equal to designing $\textbf{G$_i$},i\in[n]$.
Specifically, we first set $l=\alpha^t$.  Then we can divide the process of designing the computing scheme into two steps, including (1) designing a general message structure for $\textbf{G$_i$}$, (2) using specific segments of $s_i$ to materialize the message structure:

	(1) Designing message structure:
%As mentioned in \secref{Sec.framework}, $\textbf{G$_i$}$ is actually a combination of all segments is $s_i,i\in[n]$.
We first design a message structure of $\textbf{G$_i$}$. For the ease of description, we denote $\textbf{M}$ as a general expression for the $t$ blocks stored by each device. $\textbf{M$_u$}$ means the $u$-th block in each device and $\textbf{M$_{u,v}$}$ is the $v$-th segment in $\textbf{M$_u$}$, $u \in [t],v\in [l]$. We use the terminology $k$-$sum$, $k \in [t]$, as an expression to denote an intermediate value which is the sum of $k$ distinct segments, all of which are drawn from different blocks, {\em i.e.}, $\textbf{M$_{p_1,q_1}$}+ \cdots+ \textbf{M$_{p_k,q_k}$} $, $p_1\not= \cdots\not= p_k$. For a particular $k$, $k$-$sum$ has $\binom t k$ types, we require each type appear $(\alpha-1)^{t-k}$ times. In the message structure, we require that all types of $k$-$sum$ should be included and each segment won't be reused. For example, $\alpha=3$, $t=3$, $\textbf{M}=\{\textbf{M$_1$}, \textbf{M$_2$}, \textbf{M$_3$}\}$. Hence, 1-$sum$ has $\binom 3 1=3$ kinds, each appears $(3-1)^{3-1}=4$ times; 2-$sum$ has $\binom 3 2= 3$ kinds, each appears $(3-1)^{3-2}=2$ times; 3-$sum$ has $\binom 3 3=1$ kind, each appears $(3-1)^{3-3}=1$ times. Use $\overline{\textbf{M}}_u$ to denote a segment in $\textbf{M$_u$}$, we can get the general message structure like \tabref{tab:structure}.
\begin{table}[!h]
		\caption{Message structure}
		\center
		\begin{tabular}{|c|}
			\hline   \textbf{Structure }\\
			\hline
			 \parbox [c]{1in}{\vspace{1mm}$\overline{\textbf{M}}_1$, $\overline{\textbf{M}}_1$, $\overline{\textbf{M}}_1$, $\overline{\textbf{M}}_1$} \\
			%\hline
			\parbox [c]{1in}{\vspace{1mm}$\overline{\textbf{M}}_2$, $\overline{\textbf{M}}_2$, $\overline{\textbf{M}}_2$, $\overline{\textbf{M}}_2$} \\
			%\hline
			 \parbox [c]{1in}{\vspace{1mm}$\overline{\textbf{M}}_3$, $\overline{\textbf{M}}_3$, $\overline{\textbf{M}}_3$, $\overline{\textbf{M}}_3$\vspace{1mm}} \\
			\hline
			 \parbox [c]{1in}{\vspace{1mm}$\overline{\textbf{M}}_1+\overline{\textbf{M}}_2$, $\overline{\textbf{M}}_1+\overline{\textbf{M}}_2$}\\
			%\hline
			 \parbox [c]{1in}{\vspace{1mm}$\overline{\textbf{M}}_2+\overline{\textbf{M}}_3$, $\overline{\textbf{M}}_2+\overline{\textbf{M}}_3$}\\
			%\hline
			 \parbox [c]{1in}{\vspace{1mm}$\overline{\textbf{M}}_1+\overline{\textbf{M}}_3$, $\overline{\textbf{M}}_1+\overline{\textbf{M}}_3$\vspace{1mm}}\\
			\hline
			\parbox [c]{1in}{\vspace{1mm}$\overline{\textbf{M}}_1$+$\overline{\textbf{M}}_2$+
$\overline{\textbf{M}}_3$\vspace{1mm}} \\
			\hline	
		\end{tabular}\label{tab:structure}
	\end{table}

We next show three properties of above message structure by three lemmas.
\begin{LL}\label{Lem3_1}
		In $\textbf{G$_i$}$, each block $\textbf{B}_j$ uses $\alpha^{t-1}$ different segments, $i\in [n],j\in \widetilde{V}_i$.
	\end{LL}
	\begin{proof}
		From the message structure, we have that for a particular $k$, $k\in [t]$, all $\binom t k$ types of $k$-$sum$ are included and each type appears $(\alpha-1)^{t-k}$ times. Hence, for $\textbf{M}_u,u\in [t]$, the number of its segments appear in message structure is:
		\begin{eqnarray}\label{Eq.T1}
		\begin{aligned}
		&\sum\limits_{k=1}^{t}{\binom{t-1}{k-1} (\alpha-1)^{t-k}}=\alpha^{t-1}.
		\end{aligned}
		\end{eqnarray}
		Meanwhile, because $\textbf{B}_j$ has $l=\alpha^t>\alpha^{t-1}$ segments, so each segment won't be reused in message structure. Since $\textbf{G$_i$}$ is stemmed from the message structure, each block $\textbf{B}_j$ uses $\alpha^{t-1}$ different segments in $\textbf{G$_i$}$, $i\in [n],j\in \widetilde{V}_i$.
	\end{proof}

\begin{LL}\label{Lem3_2}
		In $\textbf{G$_i$}$, the total number of the intermediate values is $\alpha^t-(\alpha-1)^t$, {i.e.}, $g=\alpha^t-(\alpha-1)^t$.
	\end{LL}
	\begin{proof}
		From the message structure, we can easily get the total number of the intermediate values in $\textbf{G$_i$}$ is:
		\begin{eqnarray}\label{Eq.T2}
		\begin{aligned}
		&\sum\limits_{k=1}^{t}{\binom{t}{k} (\alpha-1)^{t-k}}
		%=\sum\limits_{x=0}^{t-1}{\binom{t}{x} (\alpha-1)^x}
		=\alpha^t-(\alpha-1)^t.
		\end{aligned}
		\end{eqnarray}
 Meanwhile, we have $\textbf{G$_i$}=\textbf{Q$_{i}$}\circledast\textbf{F$_i$}$. Therefore, from \ddref{Def.BMM}, we have that there are $\alpha^t-(\alpha-1)^t$ rows in $\textbf{Q$_{i}$}$, followed by $g=\alpha^t-(\alpha-1)^t$.

	\end{proof}

\begin{LL}\label{Lem3_3}
		In $\textbf{G$_i$}$, for $\forall p,q \in [t],p\not=q$, the number of the intermediate values, where the segments of $\textbf{M$_p$},\textbf{M$_q$}$ appear together, is $\alpha^{t-2}$.
	\end{LL}
	\begin{proof}
		From the message structure, we can easily get the number of the intermediate values where the segments of $\textbf{M$_p$},\textbf{M$_q$}$ appear together is:
		\begin{eqnarray}\label{Eq.T2}
		\begin{aligned}
		&\sum\limits_{k=2}^{t}{\binom{t-2}{k-2} (\alpha-1)^{t-k}}
		=\alpha^{t-2}.
		\end{aligned}
		\end{eqnarray}
	\end{proof}

\begin{RM}
The above three lemmas are useful for analyzing the property of following materialization process and proving that PCC satisfies the privacy and availability conditions.
\end{RM}

	(2) Materialization: We then use segments to materialize the message structure. Like GPC, we use the random matrix $\textbf{P}$ to select the segments randomly. Hence, the arrangement of segments in $\textbf{B$_j$}$ is equal to arranging the segments in $\textbf{B}'_j$. Firstly, we require that each segment won't be reused in one edge device. Secondly, in all edge devices, we use all $\alpha^t$ segments of $\textbf{B}'_\theta$ in an ascending order. Thirdly, for each block $\textbf{B}'_j,j\not=\theta$,
the selection of its segments has to follow three rules. First, in the $1$-$sum$ of all edge devices, we use the segments in an ascending order. Second, in one edge device, we use the segments appear with the segments of $\textbf{B}'_\theta$ in an ascending order. Third, in one edge device, we use the segments in a descending order. The three rules are from strong to weak.

We next show a property of above materialization process in \ttref{numx}.
\begin{TT}\label{numx}
%In $\textbf{R$_{i}$}$, where $s_i$ holds $\textbf{T$_\theta$}$,
For each rearranged block $\textbf{B}'_j,j\not=\theta$, the indexes of its segments that appear with the segments of $\textbf{B}'_\theta$ are no more than $\alpha^{t-2}+(\alpha-1)^{t-1}$.
	\end{TT}
	\begin{proof}
%From \llref{Lem3_1}, we have that $\textbf{B}'_j$ has $\alpha^t$ segments and we only use $\alpha^{t-1}$ segments of them in one edge devices, $j\in[w]$.
For $s_i,i\in \{\mathcal{i}|\theta\notin \widetilde{V}_\mathcal{i}\}$, there is no segment of $\textbf{B}'_\theta$ in the intermediate values. Hence, for each rearranged block $\textbf{B}'_j,j\in\widetilde{V}_i$, the indexes of its segments that appear with the segments of $\textbf{B}'_\theta$ are obviously no more than $\alpha^{t-2}+(\alpha-1)^{t-1}$.

In $s_i,i\in \{\mathcal{i}|\theta\in \widetilde{V}_\mathcal{i}\}$, the arrangement of each segment in $\textbf{B}'_j,j\in\widetilde{V}_i,j\not=\theta$, has three situations: (1) appears alone in $1$-$sum$, (2) appears with the segment of $\textbf{B}'_\theta$ in $k$-$sum$, $k\geq2$, (3) appears with the segments of the blocks except $\textbf{B}'_\theta$ in $k$-$sum$, $k\geq2$.
Because all types of $1$-$sum$ appear $(\alpha-1)^{t-1}$ times, so there are $(\alpha-1)^{t-1}$ segments in the first situation.  From \llref{Lem3_3}, we find that there are $\alpha^{t-2}$ segments in the second situation.  As depicted in the materialization process, the selection of the segments in the second situation is in an ascending order, while that in the third situation is in a descending order.

We denote the maximum index of the segments in the second situation is $z_1$.
If $z_1>\alpha^{t-2}+(\alpha-1)^{t-1}$, then before we select $\textbf{B}'_{j,z_1}$ for the second situation, there must exists a segment $\textbf{B}'_{j,z_2}$ in the third situation, where $z_2\leq\alpha^{t-2}+(\alpha-1)^{t-1}<z_1$.
Otherwise, the index of each segment in the third situation is more than $\alpha^{t-2}+(\alpha-1)^{t-1}$. In other word, when select a segment for the first two situations, we can select any segments in $\{\textbf{B}'_{j,z}|z\in[\alpha^{t-2}+(\alpha-1)^{t-1}]\}$.
Meanwhile, there are only $(\alpha-1)^{t-1}+\alpha^{t-2}$ segments in the first two situations, and we select the segments in the second situation in an ascending order. Hence, $z_1\leq\alpha^{t-2}+(\alpha-1)^{t-1}$, which contradicts with $z_1>\alpha^{t-2}+(\alpha-1)^{t-1}$.
Actually when we select $\textbf{B}'_{j,z_2}$ for the third situation, $\textbf{B}'_{j,z_1}$ is unused. However we select $z_2$, which is less than $z_1$, which contradicts with selecting segments in a descending order.
Therefore we have $z_1\leq\alpha^{t-2}+(\alpha-1)^{t-1}$. In other word, the indexes of the segments used in the second situation are no more than $\alpha^{t-2}+(\alpha-1)^{t-1}$.
	\end{proof}

\begin{RM}
\ttref{numx} is useful for selecting specific values used in the decoding process and proving that PCC satisfies the availability condition.
\end{RM}

Secondly, we give the design of decoding process.
Let $p=\alpha^{t-2}+(\alpha-1)^{t-1}$. Then we analyze the selection of values used in the decoding process.
For one thing, each segment of $\textbf{B}'_\theta$ in the intermediate values is either appearing alone or in combination with the segments from $\textbf{B$_j$}',j\not=\theta$. For another thing, from \ttref{numx}, we have that for each rearranged block $\textbf{B$_j$}',j\not=\theta$, the indexes of its segments, that appear with the segments of $\textbf{B}'_\theta$, are no more than $p$.
Therefore, we can select the intermediate values, each of which is related with $\textbf{B}'_\theta$ or is exactly a segment, whose index is no more than $p$, from $\textbf{B}'_j,j\not=\theta$.
There is a point worth noting. If there are $\alpha+1$ edge devices that store $\textbf{B}'_\theta$, then for the last edge device of them, we won't select its intermediate values related with $\textbf{B}'_\theta$ for the decoding process.
Therefore, the selected intermediate values are related with $\alpha^t$ segments in $\textbf{B}'_\theta$ and $(w-1)p$ segments from $\textbf{B}'_j,j\not=\theta$, {\em i.e.}, $h=\alpha^t+(w-1)p$.
Hence, $\textbf{x}=[\textbf{T$_{\theta,1}$};\cdots;\textbf{T$_{\theta,\alpha^t}$};
\textbf{T$_{1,\textbf{P}_{[1,1]}}$};\cdots;\textbf{T$_{1,\textbf{P}_{[1,p]}}$};
\cdots;\textbf{T$_{\theta-1,\textbf{P}_{[1,1]}}$};\cdots;\\
\textbf{T$_{\theta-1,\textbf{P}_{[1,p]}}$};
\textbf{T$_{\theta+1,\textbf{P}_{[1,1]}}$};\cdots;\textbf{T$_{\theta+1,\textbf{P}_{[1,p]}}$};\cdots;
\textbf{T$_{w,\textbf{P}_{[1,p]}}$}]$.
Finally we have $\textbf{Q$_{i}$}\in\mathbb{F}_2^{g\times lt}$, $\textbf{C}\in \mathbb{F}_2^{n\times g}$, $\textbf{D}\in\mathbb{F}_2^{h \times h}$, where $g=\alpha^{t}+(\alpha-1)^{t-1},h=\alpha^t+(w-1)p$.

	Thirdly, based on the design of above two processes, we can give the algorithm of PCC in \algref{alg2}. To better present the algorithm, we need three definitions. First, $loc()$ is a subfunction to locate $\textbf{B$_j$}$ in $\textbf{V$_i$}$ as mentioned in GPC. Second, to achieve the requirements in the materialization process, we generate two matrices $\widehat{\textbf{M}},\widetilde{\textbf{M}}\in \mathbb{F}_2^{n\times l}$. Meanwhile $\widetilde{n}()$ is a function to select an index from them as depicted in \algref{n}.
Third, we denote $\widetilde{M}_{U,k}$ as a set of all kinds of $k$-$sum$ in a set $U$. For example, if $U=\{1,2\}$, we $\widetilde{M}_{U,1}=\{1,2\}$, $\widetilde{M}_{U,2}=\{1+2\}$.
In \algref{alg2}, we first initialize the parameters and judge the solvable conditions, which will be given in \secref{proofPC} (line 1-4). If the variables don't meet solvable conditions, PCC has no solution and \algref{iPC} returns 0. Then we design for each edge device in turn. For $s_i$, we first add all kinds of $1$-$sum$ $\alpha^{t-1}$ times (line 6 to 12). For each block $\textbf{B}_j$ it stores, we select a segment from it and update $\textbf{Q$_i$}$. If $j=\theta$ or the selected index is no more than $p$, we will update $\textbf{C},\textbf{D}$.
Then we add each kind of $k$-$sum$ $\alpha^{t-k}$ times, $k\in[2:t]$ (line 13 to 23). If the type includes $\textbf{B}'_{\theta}$, we select segments and update $\textbf{Q$_i$} ,\textbf{C},\textbf{D}$. Otherwise, we select segments and update $\textbf{Q$_i$}$.
After the design for $s_i$, we reset $f_1,\widetilde{\textbf{M}}$ (line 24).
Finally we can get the whole $\textbf{Q}$, $\textbf{C}$ and $\textbf{D}$.
In each edge device, we select $t\alpha^{t-1}$ segments according to \llref{Lem3_1}. Hence, lines 5-23 of \algref{iPC} are looped at most $nt\alpha^{t-1}$ times and the complexity of line 8 is $O(l)$. Therefore, the complexity of \algref{iPC} is $O(nlt\alpha^{t-1})=O(nl\lceil\frac{w\alpha}{n}\rceil\alpha^{t-1})=O(nl\alpha^t)=O(nl^2)$.

\setlength{\textfloatsep}{0.1cm}
	\setlength{\floatsep}{0.1cm}
	\begin{algorithm}[t]\label{alg2}
		\LinesNumbered
		\begin{small}
			\KwIn{$\theta,S,\textbf{B},\{\widetilde{V}_i|i\in[n]\}$}%
			\KwOut{$\textbf{Q}$, $\textbf{C}$, $\textbf{D}$}
			Initialize $\textbf{Q$_i$}$, $\textbf{C}$, $\textbf{D}$,
$\widehat{\textbf{M}}$, $\widetilde{\textbf{M}}$ as zero matrices\;
			    $f_1=f_2=0$, $\alpha=\lfloor\frac{tn}{w}\rfloor,p=\alpha^{t-2}+(\alpha-1)^{t-1}$, $l=\alpha^t$\;
 			\uIf{!($\alpha\geq2$ and $(\alpha-1)^t\geq\alpha^{t-2}$)}{
 			    \Return 0.
 			}
			   \For{$i=1$ to $n$}{
			        \For{$j$ in $\widetilde{V}_i $}{
			             \For{$1$ to $\alpha^{t-1}$}{
 			               $a=\widetilde{n}(j,0,0)$, $f_1=f_1+1$,
    			           $\textbf{Q$_i$}_{[f_1,loc(i,j)l+\textbf{P}_{[j,a]}]}=1$\;
 			               \uIf{$j=\theta$ and $\widehat{\textbf{M}}_{j}$ hasn't be reset}{
  			                 $\textbf{C}_{[i,f_1]}=1$,
   			                 $f_2=f_2+1$,
      			             $\textbf{D}_{[f_2,a]}=1$\;}
    			           \ElseIf{$a\leq p$}{
  			                 $\textbf{C}_{[i,f_1]}=1$,
   			                 $f_2=f_2+1$,
   			                 $\textbf{D}_{[f_2,l+(j-1)p+a]}=1$\;}
   			              }
			        }

			%        \For{$k=2$ to $t$}{
			           \For{$j_1+j_2+\cdots+j_k$ in $ \widetilde{M}_{\widetilde{V}_i,k}, k\in[2:t]$}{
			             \For{$1$ to $\alpha^{t-k}$}{
 			               $f_1=f_1+1$\;
 			               \uIf{$\exists j_b=\theta$}{	
 			                 $a=\widetilde{n}(j_b,0,0)$, $f_2=f_2+1$, $\textbf{Q$_i$}_{[f_1,loc(i,\theta)l+\textbf{P}_{[\theta,a]}]}=1$\;
  			                 $\textbf{C}_{[i,f_1]}=1$,
   			                 $\textbf{D}_{[f_2,a]}=1$\;
			                 \For{$x$ in $\{j_1,\cdots,j_{b-1},j_{b+1},\cdots,j_k\}$ }{
                 $\textbf{Q$_i$}_{[f_1,loc(i,x)l+\textbf{P}_{[x,\widetilde{n}(x,1,0)]}]}=1$,
                 $\textbf{D}_{[f_2,l+(x-1)p+\widetilde{n}(x,1,0)}=1$\;}
     			             }
   			              \Else{
			                \For{$x$ in $\{j_1,\cdots,j_k\}$ }{	
                		%	   $a=\widetilde{n}(x)$\; %$\textbf{Q$_i$}(f_1,(loc(i,x)-1)l+a})=1$\;}
                $\textbf{Q$_i$}_{[f_1,loc(i,x)l+\textbf{P}_{[x,\widetilde{n}(x,1,1)]}]}=1$\;}
     			           }
			             }
			           }
			 %       }
			   $f_1=0$, reset $\widetilde{\textbf{M}}$ as zero matrix\;
			   }
 			 $\textbf{Q}\gets \bigcup_{i=1}^{n}\textbf{Q$_i$}$\;
 			\Return $\textbf{Q}$, $\textbf{C}$, $\textbf{D}$.
			\caption{PCC Algorithm } \label{iPC}
		\end{small}
	\end{algorithm}

\setlength{\textfloatsep}{0.1cm}
	\setlength{\floatsep}{0.1cm}
	\begin{algorithm}[t]%\label{alg2}
		\LinesNumbered
		\begin{small}
			\KwIn{$i,f_1,f_2$}%
			\KwOut{$j$}

 			 \uIf{$f_1=0$}{
                \uIf{$\forall j\in[l],\widehat{\textbf{M}}_{[i,j]}=1$}{
                 reset $\widehat{\textbf{M}}_{i}$ as zero vector\;
                 }

                    $j=\min\{\mathcal{j}|\widehat{\textbf{M}}_{[i,\mathcal{j}]}=0\}$, $\widehat{\textbf{M}}_{[i,j]}=1,\widetilde{\textbf{M}}_{[i,j]}=1$\;
      		 }
      		 \Else{
 			      \uIf{$f_2=0$}{
                    $j=\min \{\mathcal{j}|\widetilde{\textbf{M}}_{[i,\mathcal{j}]}=0\}$\;}
      		      \Else{
                    $j=\max\{\mathcal{j}|\widetilde{\textbf{M}}_{[i,\mathcal{j}]}=0\}$\;}
			      $\widetilde{\textbf{M}}_{[i,j]}=1$\;
      		 }

			\Return $j$.
			\caption{$\widetilde{n}$} \label{n}
		\end{small}
	\end{algorithm}

Then we take $n=4, w=4, t=3, \theta=1$ for example and show the computing schemes of GPC, PCC respectively. There are $n=4$ edge devices and $w=4$ blocks. Each edge device stores $t=3$ blocks and each block appears in $\alpha=\lfloor\frac{tn}{w}\rfloor=3$ edge devices.
Firstly, we show the scheme of GPC in \tabref{tab:examples_gpc}.
Each block has $l=\alpha=3$ segments and $\textbf{B}'_1$ uses three segments. Hence, the user can recover $\textbf{T}_1$ after $s_i$ computes $\textbf{A}\times\textbf{G$_i$}$ and returns the answer, $i\in[n]$.
For $s_1$, we can easily find that each block it stores uses one segment.
Meanwhile since $\textbf{B}'_{j,v}=\textbf{B}_{j,\textbf{P}_{[j,v]}}$ and $\textbf{$\textbf{P}$}$ is a random matrix, $s_1$ cannot infer anything from the indexes of segments. Hence, $s_1$ cannot identify the target block and it's the same to other devices.
Secondly, we show the scheme of PCC in \tabref{tab:examples}. Each block has $l=\alpha^t=27$ segments and $\textbf{B}'_1$ uses all segments. Meanwhile, each segment of $\textbf{B}'_1$ is either appearing alone or in combination with the segments appearing in $1$-$sum$. Hence, the user can recover $\textbf{T}_1$ after $s_i$ computes $\textbf{A}\times\textbf{G$_i$}$ and returns the answer, $i\in[n]$.
For $s_1$, we can easily find that each block it stores uses nine different segments. Meanwhile like GPC, $s_1$ cannot infer anything from the indexes of segments. Hence, $s_1$ cannot identify the target block and it's the same to other devices.

\begin{table}\
		\caption{An example of GPC where $\theta=1$}
		\center
		\begin{tabular}{|c|c|c|c|}
			\hline   			
 			 \parbox [c]{0.7in}{\vspace{1mm}$s_1(\textbf{B}_1,\textbf{B}_2,\textbf{B}_3)$}&
			 \parbox [c]{0.7in}{\vspace{1mm}$s_2(\textbf{B}_1,\textbf{B}_2,\textbf{B}_4)$}&
			 \parbox [c]{0.7in}{\vspace{1mm}$s_3(\textbf{B}_1,\textbf{B}_3,\textbf{B}_4)$}&
			 \parbox [c]{0.7in}{\vspace{1mm}$s_4(\textbf{B}_2,\textbf{B}_3,\textbf{B}_4)$}\\
			\hline
			 \parbox [c]{0.3in}{\vspace{1mm}$\textbf{B}'_{1,1}$}&
			 \parbox [c]{0.3in}{\vspace{1mm}$\textbf{B}'_{1,2}$}&
			 \parbox [c]{0.3in}{\vspace{1mm}$\textbf{B}'_{1,3}$}&
			 \parbox [c]{0.3in}{\vspace{1mm}$\textbf{B}'_{2,1}$}\\

			\hline
			 \parbox [c]{0.3in}{\vspace{1mm}$\textbf{B}'_{2,1}$}&
			 \parbox [c]{0.3in}{\vspace{1mm}$\textbf{B}'_{2,1}$}&
			 \parbox [c]{0.3in}{\vspace{1mm}$\textbf{B}'_{3,1}$}&
			 \parbox [c]{0.3in}{\vspace{1mm}$\textbf{B}'_{3,1}$}\\

			\hline
			 \parbox [c]{0.3in}{\vspace{1mm}$\textbf{B}'_{3,1}$}&
			 \parbox [c]{0.3in}{\vspace{1mm}$\textbf{B}'_{4,1}$}&
			 \parbox [c]{0.3in}{\vspace{1mm}$\textbf{B}'_{4,1}$}&
			 \parbox [c]{0.3in}{\vspace{1mm}$\textbf{B}'_{4,1}$}\\
			\hline	
		\end{tabular}\label{tab:examples_gpc}
	\end{table}

\begin{table*}\
		\caption{An example of PCC where $\theta=1$}
		\center
		\begin{tabular}{|l|l|l|l|}
			\hline   			
 			 \parbox [c]{1.4in}{\vspace{1mm}$s_1(\textbf{B}_1,\textbf{B}_2,\textbf{B}_3)$}&
			 \parbox [c]{1.4in}{\vspace{1mm}$s_2(\textbf{B}_1,\textbf{B}_2,\textbf{B}_4)$}&
			 \parbox [c]{1.4in}{\vspace{1mm}$s_3(\textbf{B}_1,\textbf{B}_3,\textbf{B}_4)$}&
			 \parbox [c]{1.4in}{\vspace{1mm}$s_4(\textbf{B}_2,\textbf{B}_3,\textbf{B}_4)$}\\
			\hline
			 \parbox [c]{1.4in}{\vspace{1mm}$\textbf{B}'_{1,1}$, $\textbf{B}'_{1,2}$, $\textbf{B}'_{1,3}$, $\textbf{B}'_{1,4}$}&
			 \parbox [c]{1.4in}{$\textbf{B}'_{1,10}$, $\textbf{B}'_{1,11}$, $\textbf{B}'_{1,12}$, $\textbf{B}'_{1,13}$}&
			\parbox [c]{1.4in}{$\textbf{B}'_{1,19}$, $\textbf{B}'_{1,20}$, $\textbf{B}'_{1,21}$, $\textbf{B}'_{1,22}$}&
			 \parbox [c]{1.4in}{$\textbf{B}'_{2,9}$, $\textbf{B}'_{2,10}$, $\textbf{B}'_{2,11}$, $\textbf{B}'_{2,12}$}\\
			
			 \parbox [c]{1.4in}{$\textbf{B}'_{2,1}$, $\textbf{B}'_{2,2}$, $\textbf{B}'_{2,3}$, $\textbf{B}'_{2,4}$}&
			 \parbox [c]{1.4in}{$\textbf{B}'_{2,5}$,  $\textbf{B}'_{2,6}$,  $\textbf{B}'_{2,7}$,  $\textbf{B}'_{2,8}$}&
			 \parbox [c]{1.4in}{$\textbf{B}'_{3,5}$, $\textbf{B}'_{3,6}$,  $\textbf{B}'_{3,7}$,  $\textbf{B}'_{3,8}$}&
			 \parbox [c]{1.4in}{$\textbf{B}'_{3,9}$, $\textbf{B}'_{3,10}$,  $\textbf{B}'_{3,11}$,  $\textbf{B}'_{3,12}$}\\

			 \parbox [c]{1.4in}{$\textbf{B}'_{3,1}$, $\textbf{B}'_{3,2}$, $\textbf{B}'_{3,3}$, $\textbf{B}'_{3,4}$\vspace{1mm}}&
			 \parbox [c]{1.4in}{$\textbf{B}'_{4,1}$, $\textbf{B}'_{4,2}$, $\textbf{B}'_{4,3}$, $\textbf{B}'_{4,4}$\vspace{1mm}}&
			 \parbox [c]{1.4in}{$\textbf{B}'_{4,5}$, $\textbf{B}'_{4,6}$, $\textbf{B}'_{4,7}$, $\textbf{B}'_{4,8}$\vspace{1mm}}&
			 \parbox [c]{1.4in}{$\textbf{B}'_{4,9}$, $\textbf{B}'_{4,10}$, $\textbf{B}'_{4,11}$, $\textbf{B}'_{4,12}$\vspace{1mm}}\\

			\hline
			 \parbox [c]{1.35in}{\vspace{1mm}$\textbf{B}'_{1,5}+\textbf{B}'_{2,5}$, $\textbf{B}'_{1,6}+\textbf{B}'_{2,6}$}&
			 \parbox [c]{1.35in}{\vspace{1mm}$\textbf{B}'_{1,14}+\textbf{B}'_{2,1}$, $\textbf{B}'_{1,15}+\textbf{B}'_{2,2}$}&
			 \parbox [c]{1.35in}{\vspace{1mm}$\textbf{B}'_{1,23}+\textbf{B}'_{3,1}$, $\textbf{B}'_{1,24}+\textbf{B}'_{3,2}$}&
			 \parbox [c]{1.35in}{\vspace{1mm}$\textbf{B}'_{2,27}+\textbf{B}'_{3,27}$, $\textbf{B}'_{2,26}+\textbf{B}'_{3,26}$}\\

			%\hline
			 \parbox [c]{1.35in}{$\textbf{B}'_{2,27}+\textbf{B}'_{3,27}$, $\textbf{B}'_{2,26}+\textbf{B}'_{3,26}$}&
			 \parbox [c]{1.35in}{$\textbf{B}'_{2,27}+\textbf{B}'_{4,27}$, $\textbf{B}'_{2,26}+\textbf{B}'_{4,26}$}&
			 \parbox [c]{1.35in}{$\textbf{B}'_{3,27}+\textbf{B}'_{4,27}$, $\textbf{B}'_{3,26}+\textbf{B}'_{4,26}$}&
			 \parbox [c]{1.35in}{$\textbf{B}'_{2,25}+\textbf{B}'_{4,27}$, $\textbf{B}'_{2,24}+\textbf{B}'_{4,26}$}\\
			%\hline
			 \parbox [c]{1.35in}{$\textbf{B}'_{1,7}+\textbf{B}'_{3,5}$, $\textbf{B}'_{1,8}+\textbf{B}'_{3,6}$\vspace{1mm}}&
			 \parbox [c]{1.35in}{$\textbf{B}'_{1,16}+\textbf{B}'_{4,5}$, $\textbf{B}'_{1,17}+\textbf{B}'_{4,6}$\vspace{1mm}}&
			 \parbox [c]{1.35in}{$\textbf{B}'_{1,25}+\textbf{B}'_{4,1}$, $\textbf{B}'_{1,26}+\textbf{B}'_{4,2}$\vspace{1mm}}&
			 \parbox [c]{1.35in}{$\textbf{B}'_{3,25}+\textbf{B}'_{4,25}$, $\textbf{B}'_{3,24}+\textbf{B}'_{4,24}$\vspace{1mm}}\\
			\hline
			 \parbox [c]{1.1in}{\vspace{1mm}$\textbf{B}'_{1,9}+\textbf{B}'_{2,7}+\textbf{B}'_{3,7}$\vspace{1mm}}&
			 \parbox [c]{1.1in}{\vspace{1mm}$\textbf{B}'_{1,18}+\textbf{B}'_{2,3}+\textbf{B}'_{4,7}$\vspace{1mm}}&
			 \parbox [c]{1.0in}{\vspace{1mm}$\textbf{B}'_{1,27}+\textbf{B}'_{3,3}+\textbf{B}'_{4,3}$\vspace{1mm}}&
			 \parbox [c]{1.0in}{\vspace{1mm}$\textbf{B}'_{2,23}+\textbf{B}'_{3,23}+\textbf{B}'_{4,23}$\vspace{1mm}}\\
			\hline	
		\end{tabular}\label{tab:examples}
	\end{table*}

	\section{Theoretical analysis}\label{TA}
In this section, we give theoretical analysis on the proposed computing schemes. Firstly, we prove that they satisfy the privacy and availability conditions. Secondly, we analyze their storage load, communication load and computing complex in each edge device. Thirdly, we compare them with other private schemes.

\subsection{Proof of Privacy and Availability Conditions}\label{proofPC}
		Firstly, we prove that GPC satisfies the privacy and availability conditions.
\begin{TT}\label{PrivayProof}
		GPC satisfies the privacy and availability conditions.
	\end{TT}
	\begin{proof}
		Firstly, we prove that GPC satisfies the privacy condition. By the chain rule, we can write the privacy constraint in \ddref{Def.Privacy} as follows:
	    \begin{eqnarray}\label{Eq.privacychain}
		\begin{aligned} 		
	&\textbf{I}(\theta;\textbf{Q$_{i}$},\textbf{A},\textbf{R$_{i}$},\textbf{V$_{i}$})
		=\textbf{I}(\theta;\textbf{Q$_{i}$})
		+\textbf{I}(\theta;\textbf{V$_{i}$}|\textbf{Q$_{i}$})\\
		+&\textbf{I}(\theta;\textbf{A}|\textbf{Q$_{i}$},\textbf{V$_{i}$})
	+\textbf{I}(\theta;\textbf{R$_{i}$}|\textbf{Q$_{i}$},\textbf{V$_{i}$},\textbf{A}).
		\end{aligned}
		\end{eqnarray}

		Since \textbf{A} is a local data of user and has nothing with $\theta$, $\textbf{I}(\theta;\textbf{A}|\textbf{Q$_{i}$},\textbf{V$_{i}$})=0$. Meanwhile $\textbf{V$_{i}$}$ and $\textbf{F$_{i}$}$ are stemmed from the storage allocation phase and the cloud determines the storage allocation without knowing any information of $\theta$. Hence, $\textbf{V$_{i}$}$ and $\textbf{F$_{i}$}$ are independent of $\theta$, which is followed by $\textbf{I}(\theta;\textbf{V$_{i}$}|\textbf{Q$_{i}$})=0$.

For each block $\textbf{B}_{i_b}$ in $s_i$, we select one segment from it and denote the index of this segment as $o_b,b\in[t]$. Assuming the above block is $\textbf{B$_j$}, j\in[w]$, we have $o_b=\textbf{P}_{[j,k]},k\in[\alpha]$, which is actually an element of $\textbf{P}$. Since $\textbf{P}$ is a random matrix, $o_1,\cdots,o_t$ are chosen independently of each other and also independently of $\theta$. Because we select and only select one segment from $\textbf{M$_{b}$}$ in GPC, $\textbf{Q$_i$}$ is completely determined by $\{o_1,\cdots,o_t\}$.
Thus $\textbf{Q$_i$}$ is not depend on $\theta$, which implies $\textbf{I}(\theta;\textbf{Q$_i$})=0$.
Note that \textbf{R$_{i}$} is determined by $\textbf{A},\textbf{F$_{i}$},\textbf{Q$_i$}$ and all of them are independent of $\theta$. Hence we have $\textbf{I}(\theta;\textbf{R$_{i}$}|\textbf{Q$_i$},\textbf{V$_{i}$},\textbf{A})=0$.
Therefore, $\textbf{I}(\theta;\textbf{Q$_i$},\textbf{A},\textbf{R$_{i}$},\textbf{V$_{i}$})=0,i\in[n]$. It means GPC satisfies the privacy condition.

		Secondly, we prove that GPC satisfies the availability condition.
For one thing, since $\textbf{D}\in\mathbb{F}_2^{h \times h}, h=\alpha\geq1$ in GPC, we have $Rank(\textbf{D})\leq \alpha$.
For another thing, we select $\alpha$ values for the decoding process and each of them is exactly a segment in $\textbf{x}=[\textbf{T$_{\theta,1}$},\cdots,\textbf{T$_{\theta,\alpha}$}]$.
Since the $\alpha$ values are different, the row vectors of them in $\textbf{D}$ are linearly independent, {\em i.e.}, $Rank(\textbf{D})\geq \alpha$. Therefore, we have $Rank(\textbf{D})=\alpha=h$. It means GPC satisfies the availability condition.
	\end{proof}

\ttref{PrivayProof} shows that GPC can be applied in general case. Next we prove that PCC satisfy the privacy and availability conditions.

	\begin{TT}\label{PrivayProof_2}
	 If $(\alpha-1)^t\geq \alpha^{t-2}$ and $\alpha=\lfloor\frac{tn}{w}\rfloor\geq2$, PCC satisfies the privacy and availability conditions.
	\end{TT}
	\begin{proof}
		Firstly, we prove that PCC satisfies the privacy condition.
 As analyzed in \ttref{PrivayProof}, we have
$\textbf{I}(\theta;\textbf{A}|\textbf{Q$_{i}$},\textbf{V$_{i}$})=0$ and $\textbf{I}(\theta;\textbf{V$_{i}$}|\textbf{Q$_{i}$})=0$.
From the design of PCC, we have that $\textbf{Q$_{i}$}$ is completely determined by message structure and the materialization process.
For one thing, the message structure is fixed regardless of $\theta$ as designed in \secref{IPC}. For another thing, the materialization process is equal to determining the order of segments from each block. For each block $\textbf{B}_{i_b}$ in $s_i$, we use $\alpha^{t-1}$ different segments, $b\in[t]$. Then we denote the order of its used segments as $\bar{o_b}=[o_{b,1}, \cdots, o_{b,\alpha^{t-1}}]$. As analyzed in \ttref{PrivayProof}, $o_{b,u}$ is actually a element of $\textbf{P}$, $u\in[\alpha^{t-1}]$. Since $\textbf{P}$ is a random matrix, $\bar{o}_{{i_1}},\cdots,\bar{o}_{{i_t}}$ are chosen independently of each other and also independently of $\theta$.
Therefore, $\textbf{Q$_{i}$}$ is not depend on $\theta$, which implies $\textbf{I}(\theta;\textbf{Q$_{i}$})=0$. %Since $Q$ is the set of all query messages, so we have that $\textbf{I}(\theta;|Q_j)=0$.
		 Note that \textbf{R$_{i}$} is determined by $\textbf{A},\textbf{F$_{i}$},\textbf{Q$_{i}$}$ and all of them are independent of $\theta$. Hence, we have $\textbf{I}(\theta;\textbf{R$_{i}$}|\textbf{Q$_{i}$},\textbf{V$_{i}$},\textbf{A})=0$.
		Therefore, $\textbf{I}(\theta;\textbf{Q$_i$},\textbf{A},\textbf{R$_{i}$},\textbf{V$_{i}$})=0,i\in[n]$. It means PCC satisfies the privacy condition.

		 Next we prove that if $(\alpha-1)^t\geq \alpha^{t-2}$ and $\alpha=\lfloor\frac{tn}{w}\rfloor\geq2$, PCC satisfies the availability condition.
Firstly, as mentioned in message structure, all types of $1$-$sum$ are consist of one segment and appear $(\alpha-1)^{t-1}$ times. It means for each block $s_i$ stores, we can directly get $(\alpha-1)^{t-1}$ segments of it. Secondly, $\textbf{B$_j$}$ appears in at least $\alpha$ edge devices, $j\in[w]$. Hence, we can directly get at least $\alpha(\alpha-1)^{t-1}$ segments of $\textbf{B$_j$}$ from $1$-$sum$.
Thirdly, when selecting segments from $\textbf{B$_j$}, j\not=\theta$ in the $1$-$sum$ of all edge devices, we use the segments of $\textbf{B}'_j$ in an ascending order.
Therefore, for $\textbf{B$_j$}, j\not=\theta$, we can directly get $\{\textbf{B}_{j,\textbf{P}_{[1,v]}}|v\in[\alpha(\alpha-1)^{t-1}]\}$ %$\textbf{B}_{j,\textbf{P}_{[1,1]}},\cdots,\textbf{B$_{j,\textbf{P}_{[1,\alpha(\alpha-1)^{t-1}]}}$}$
from $1$-$sum$ of $\{\textbf{G}_i|i\in[n]\}$.
Hence, after every edge device $s_i$ computes $\textbf{R$_i$}=\textbf{A}\times\textbf{G$_i$}$ and returns it to $s_0$, we can directly get
$\{\textbf{T}_{j,\textbf{P}_{[1,v]}}|v\in[\alpha(\alpha-1)^{t-1}]\},j\not=\theta$.
If $(\alpha-1)^t\geq \alpha^{t-2},\alpha\geq2$, we have $\alpha(\alpha-1)^{t-1}=(\alpha-1)(\alpha-1)^{t-1}+ (\alpha-1)^{t-1}\geq \alpha^{t-2}+(\alpha-1)^{t-1}=p$. In other word, for $\textbf{T$_j$}, j\not=\theta$, we can directly get %$\textbf{T$_{j,\textbf{P}_{[1,1]}}$},\cdots,\textbf{T$_{j,\textbf{P}_{[1,p]}}$}$
$\{\textbf{T}_{j,\textbf{P}_{[1,v]}}|v\in[p]\}$
from $1$-$sum$ of $\textbf{R}$.
We select these $p(w-1)$ values for the decoding process and each of them is exactly a segment in $\textbf{x}$.
Since the $p(w-1)$ values are different, the row vectors of them in $\textbf{D}$ are linearly independent.

Meanwhile, for each segment of $\textbf{B$_\theta$}$, it is either appearing alone or in combination with the segments from $\textbf{B$_j$}, j\not=\theta$ in \textbf{G$_i$}. Hence, after $s_i$ computes $\textbf{R$_i$}=\textbf{A}\times\textbf{G$_i$}$, there are $\alpha$ values related with $\textbf{T$_\theta$}$ and the row vectors of them in $\textbf{D}$ are linearly independent. Moreover, since these $\alpha$ values are related with $\textbf{T$_\theta$}$ while the above $p(w-1)$ values are not, all the row vectors of them in $\textbf{D}$ are linearly independent. Hence, $Rank(\textbf{D})\geq \alpha+p(w-1)$. Since $\textbf{D}\in\mathbb{F}_2^{h \times h},h=l+(w-1)p$, we have $Rank(\textbf{D})\leq l+(w-1)p$. Hence, $Rank(\textbf{D})=l+(w-1)p=h$. It means PCC satisfies the availability condition if $(\alpha-1)^t\geq \alpha^{t-2}$ and $\alpha=\lfloor\frac{tn}{w}\rfloor\geq2$.
	\end{proof}

\ttref{PrivayProof_2} shows that PCC can only be applied in special cases where $(\alpha-1)^t\geq \alpha^{t-2}$ and $\alpha=\lfloor\frac{tn}{w}\rfloor\geq2$.

\subsection{Resource Consumption Analysis}\label{FRT}
	In this subsection, we present the storage load, communication load and computing complex of the proposed computing schemes respectively.
\begin{TT}\label{eff_1}
		Given the storage factor $t$, GPC achieves the storage load of $t$, the communication load of $\frac{tn}{\alpha}$ and the computing complex in each edge devices is $O(rst^2+mrs)$, $\alpha=\lfloor\frac{tn}{w}\rfloor$.
	\end{TT}
	\begin{proof}
		Firstly, the storage load $s_d=t$ as defined in \secref{SA}.

		Secondly, we analyze the communication load. For one thing, each edge device $s_i$ stores $t$ blocks and splits each of them to $\alpha$ segments, $i\in[n]$.
For another thing, $s_i$ selects one segment from each block as an intermediate value.
Hence, $\textbf{R$_i$}=\textbf{A}\times\textbf{G$_i$}$ is consist of $t$ values and the size of each is $\frac{1}{\alpha}$ of $\textbf{T}_\theta$. Therefore, GPC achieves a communication load of $c_d=\frac{tn}{\alpha}$. When $tn$ is divisible by $w$, $c_d=w$. Otherwise $c_d=\lceil\frac{\alpha w}{n}\rceil* \frac{n}{\alpha}>\frac{\alpha w}{n}*\frac{n}{\alpha}=w$. In general, the communication load fluctuates around $w$.

		Thirdly, we analyze the computing complex. For one thing, $s_i$ computes $\textbf{G$_{i}$}=\textbf{Q$_{i}$}\circledast \textbf{F$_i$}$ where $\textbf{Q$_{i}$} \in \mathbb{F}_2^{g\times lt}$. Meanwhile $\textbf{F$_i$}$ has $lt$ segments and each of them has a dimension of $r\times\frac{s}{l}$. Hence, the computing complex of this step is $O(gtrs)$.
For another thing, $s_i$ computes $\textbf{R$_{i}$}=\textbf{A}\times\textbf{G$_i$}$, where $\textbf{A}\in \mathbb{F}_q^{m\times r}, \textbf{G$_i$}\in\mathbb{F}_q^{r\times gs/l}$. The computing complex of this step is $O(mrgs/l)$.
The computing complex of the above two steps is $O(gtrs+mrgs/l)$. Since $g=t,l=\alpha=\lfloor\frac{tn}{w}\rfloor$ in GPC, $O(rst^2+mrst/\lfloor\frac{tn}{w}\rfloor)=O(rst^2+mrs)$. %Therefore, the computing complex in edge devices is $O(nrst^2+nmrs)$.
It is obviously that computing complex increases with the increase of $t$.
	\end{proof}

\iffalse
For given $\alpha \in[n]$, we denote $\beta$ as $\{\mathcal{t}|\lfloor\frac{\mathcal{t}n}{w}\rfloor=\alpha,\mathcal{t}\in\mathbb{Z^+}\}$. Then we have the following corollary.
\begin{CC}\label{load_gpc}
		For given $\alpha\in[n]$, the communication load in GPC increases with the increase of the storage load $t$, where $t\in\beta$.
	\end{CC}
	\begin{proof}
For one thing, GPC achieves a communication load of $c_d=\frac{tn}{\alpha}$.
For another thing, for $t\in\beta$, we can get the same value of $\alpha$.
Hence, for $t\in\beta$, we can take $\alpha$ in $c_d=\frac{tn}{\alpha}$ as a constant.
Then we can easily find that $c_d=\frac{tn}{\alpha}$ increases with the increase of $t$.
Therefore, for given $\alpha\in[n]$, the communication load in GPC increases with the increase of $t$, where $t\in\beta$.
	\end{proof}

\begin{RM}
		Although in general, the communication load of GPC fluctuates around $w$. But \ccref{load_gpc} shows that sometimes, the communication load in GPC may increase with the increase of storage load $t$. It's due to the existence of the rounding function.
\end{RM}

\fi

\begin{TT}\label{eff_2}
		Given the storage factor $t$, PCC achieves the storage load of $t$, the communication load of $n[1-(1-\frac{1}{\alpha})^t]$ and the computing complex in each edge device is $O(rstg+mrs),g=\alpha^t-(\alpha-1)^t$, $\alpha=\lfloor\frac{tn}{w}\rfloor$.
	\end{TT}
	\begin{proof}
				Firstly, the storage load $s_d=t$ as defined in \secref{SA}.

		Secondly, we analyze the communication load. For one thing, each edge device $s_i$ stores $t$ blocks and splits each of them to $l=\alpha^t$ segments to get $\textbf{F$_i$}$, $i\in[n]$. Since $\textbf{G$_{i}$}=\textbf{Q$_{i}$}\circledast \textbf{F$_i$}$, from the definition of $\circledast$, we have that the size of the intermediate values in $\textbf{G$_{i}$}$ is the same with that of each segment, {\em i.e.}, $\frac{1}{\alpha^t}$ of $\textbf{B}_\theta$.
%Meanwhile as shown in \ddref{Def.BMM}, the size of the intermediate value is the same with the size of each segment, {\em i.e.}, $\frac{1}{\alpha^t}$ of $\textbf{B}_\theta$.
For another thing, from \llref{Lem3_2}, we have that the total number of the intermediate values in $\textbf{G$_{i}$}$ is $\alpha^t-(\alpha-1)^t$.
Hence, $\textbf{R$_i$}=\textbf{A}\times\textbf{G$_i$}$ is consist of $\alpha^t-(\alpha-1)^t$ values and the size of each is $\frac{1}{\alpha^t}$ of $\textbf{T}_\theta$. In other word, PCC achieves a communication load of
	    \begin{eqnarray}\label{Eq.T4}
		\begin{aligned}
		 c_{d}=&\sum\limits_{j=1}^n c_j
		 %=&\frac{n}{L}[\alpha^t-(\alpha-1)^t]\\
		 =\frac{n}{\alpha^t}[\alpha^t-(\alpha-1)^t]
		 %=&n[1-(\frac{\alpha-1}{\alpha})^t]\\
		 =n[1-(1-\frac{1}{\alpha})^t].
		\end{aligned}
		\end{eqnarray}

		Thirdly,  we analyze the computing complex. As mentioned in \ttref{eff_1}, $s_i$ computes $\textbf{G$_{i}$}=\textbf{Q$_{i}$}\circledast \textbf{F$_i$}$ and $\textbf{R$_{i}$}=\textbf{A}\times\textbf{G$_i$}$.
The computing complex of the two steps is $O(gtrs+mrgs/l)$. The difference is $g=\alpha^t-(\alpha-1)^t,l=\alpha^t$ in PCC. Therefore, PCC achieves a computing complex of $O(rstg+mrs),g=\alpha^t-(\alpha-1)^t$.
	\end{proof}

\begin{RM}
		Since $\textbf{Q$_{i}$}\in\mathbb{F}_2$ and it has many $0$ elements, the actual computing quantity of the first step is not as high as analyzed.
Each edge device gets $\alpha^t-(\alpha-1)^t$ intermediate values from $t\alpha^{t-1}$ segments. Hence, the total number of additions is $t\alpha^{t-1}-\alpha^t+(\alpha-1)^t$. Meanwhile, each segment is with a dimension of $r\times \frac{s}{\alpha^t}$. Therefore the computing complexity of the additions is $O((t\alpha^{t-1}-\alpha^t+(\alpha-1)^t)\frac{rs}{\alpha^t})=O(rs)$.
\end{RM}

		\subsection{Comparison}\label{EDC}
In this subsection, we compare the computing complex, communication load, and applicability of RPC, VPC, GPC, PCC and the following two private schemes:
	\begin{itemize}
		\item
		\emph{Polynomial code based Private Computing} (PPC) scheme. In \cite{M1} where each devices share a public library, the author use polynomial code to protect the security of local data and the user's privacy. Here we only consider user's privacy and modify the original scheme as a comparison scheme. Specifically, we divide each blocks into $l$ segments and use polynomial code to encode them, where $l$ is a variable which we can adjust randomly.
		\item
		\emph{PIR based Private Computing} (PIRPC) scheme. In \cite{M9} where each devices share a public library, the author propose the PIR scheme which is of good symmetry in structure to complete the private information retrieval. Here we use the original PIR scheme to complete the matrix multiplication tasks.
	\end{itemize}
Since the cloud has powerful computing capacity, we focus on the computing process in each edge device. It's worth noting that we only consider $n,w,t\geq2$ and $\alpha=\lfloor\frac{tn}{w}\rfloor\geq1$ as analyzed in \secref{SA}. The related information is summarized in \tabref{tab:complex}.

From \tabref{tab:complex}, we find that
(1) in terms of computing complex:  PIRPC, PCC$>$GPC, RPC, VPC, PPC;
(2) in terms of communication load: RPC$>$VPC,GPC$>$PCC$>$PIRPC, PPC;
(3) in terms of applicability: GPC, RPC$>$PCC$>$VPC$>$PIRPC, PPC.
Only RPC, VPC, GPC and PCC can be applied in PEC problem where edge devices may store part of the library. With similar computing complex, GPC outperforms RPC, VPC in terms of communication load and applicability. Although having higher computing complex, PCC can further reduce the communication load.

\begin{table*}[htbp]%*意思是框两行
		\caption{Performance comparison of different schemes.}
		\center
		\begin{tabular}{|c|c|c|c|}
			\hline \begin{minipage}{1.3cm}\vspace{2mm}  \vspace{2mm} \end{minipage} & \textbf{computing complex}& \textbf{communication load}& \textbf{solvable conditions}\\
			\hline
			\begin{minipage}{1.3cm}\vspace{2mm} PPC\vspace{2mm} \end{minipage}& $O(rst+mrs/l)$ &$\frac{n}{l}$ & $l\leq n-1,t=w$\\

			\hline
			\begin{minipage}{1.3cm}\vspace{2mm} PIRPC\vspace{2mm} \end{minipage}& \tabincell{c}{$O(rstg+mrs)$,\\$g=\alpha^{t-1}-\frac{\alpha^{t-1}-1}{\alpha-1}$} & $1+\frac{1}{\alpha}+\cdots\frac{1}{\alpha^{t-1}}$ & $t=w$ \\	

			\hline
			\begin{minipage}{1.3cm}\vspace{2mm} RPC\vspace{2mm} \end{minipage}& $O(mrst)$ & $nt$ &  \\	
			\hline

			\begin{minipage}{1.3cm}\vspace{2mm} VPC\vspace{2mm} \end{minipage}& $O(rst+mrs)$ &$n$ & $n\geq w$ \\
			\hline
			\begin{minipage}{1.3cm}\vspace{2mm} GPC\vspace{2mm} \end{minipage}& $O(rst^2+mrs)$  &$\frac{nt}{\alpha}$ &  \\
			\hline
			\begin{minipage}{1.3cm}\vspace{3mm} IPC\vspace{3mm} \end{minipage}& \tabincell{c}{$O(rstg+mrs)$,\\$g=\alpha^t-(\alpha-1)^t$} & $n[1-(1-\frac{1}{\alpha})^t]$ & \tabincell{c}{ $\alpha\geq2$,\\$(\alpha-1)^t\geq\alpha^{t-2}$ } \\	
			\hline	
		\end{tabular}\label{tab:complex}
\vspace{0.5cm}
	\end{table*}

	\section{Numerical Experiments}
	\label{sec.sim}
	In this section, we conduct extensive simulation experiments to evaluate the performance of the proposed computing schemes in PEC problem. Specifically, we compare GPC, VPC, RPC and PCC. In the simulation, we consider three parameters: (1) $w$, which is the number of blocks in $\textbf{B}$, (2) $n$, which is the number of edge devices, (3) $t_0$, which is storage limit of each edge device. As defined in \secref{sec.sm}, the communication load is measured by the number of elements in $\textbf{R$_{i}$}$ and normalized by the size of $\textbf{T$_\theta$}$, {\em i.e.}, it's independent of $m,r,s$. Hence, we won't take $m,r,s$ as parameters in the simulation.

\begin{figure*}[htbp] \centering
\begin{minipage}[t]{0.2\linewidth}
\centering
\includegraphics[width=1.45in]{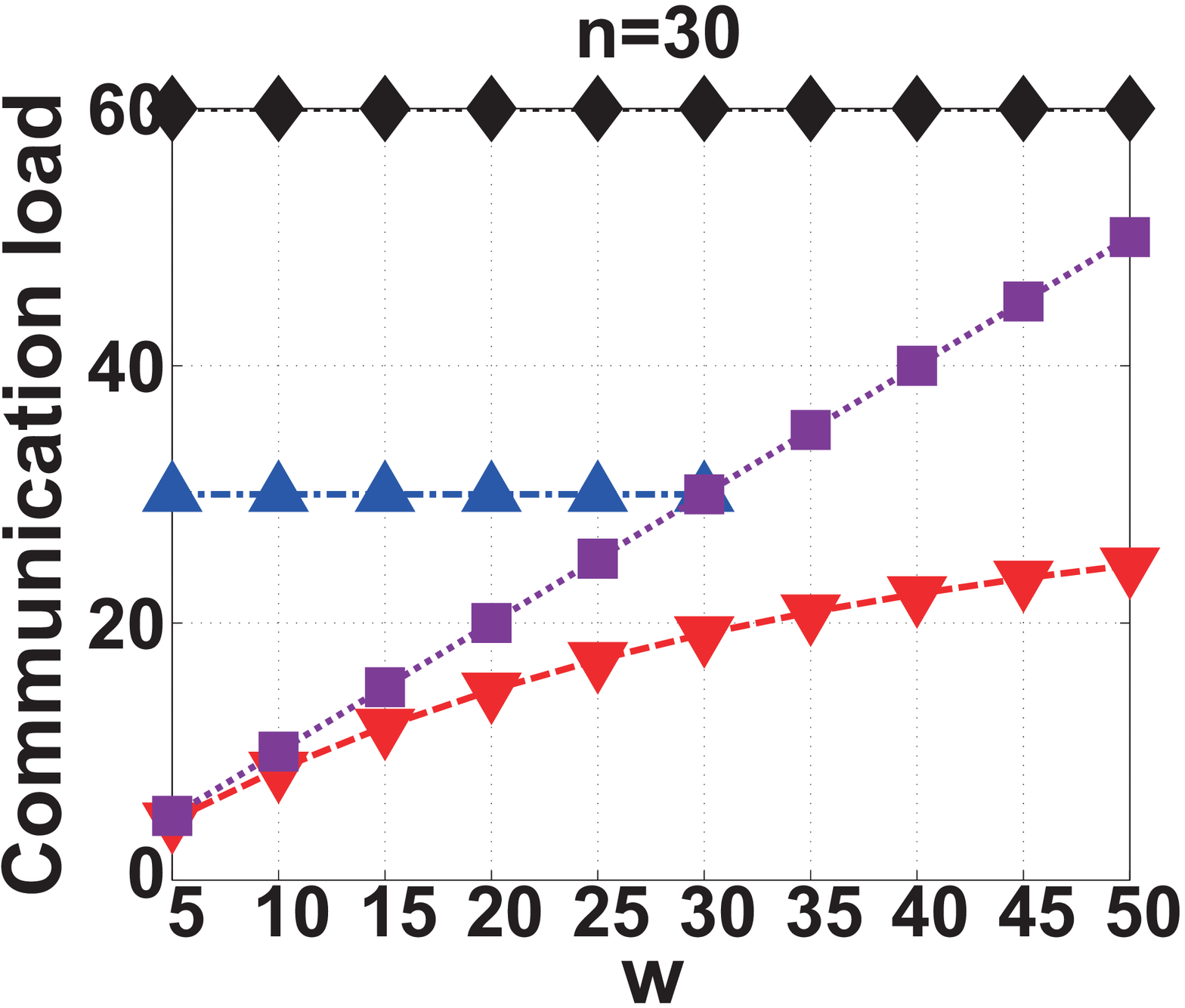}
%\mbox{\footnotesize (a) The private polynomial code.}
\end{minipage}%
\begin{minipage}[t]{0.2\linewidth}
\centering
\includegraphics[width=1.45in]{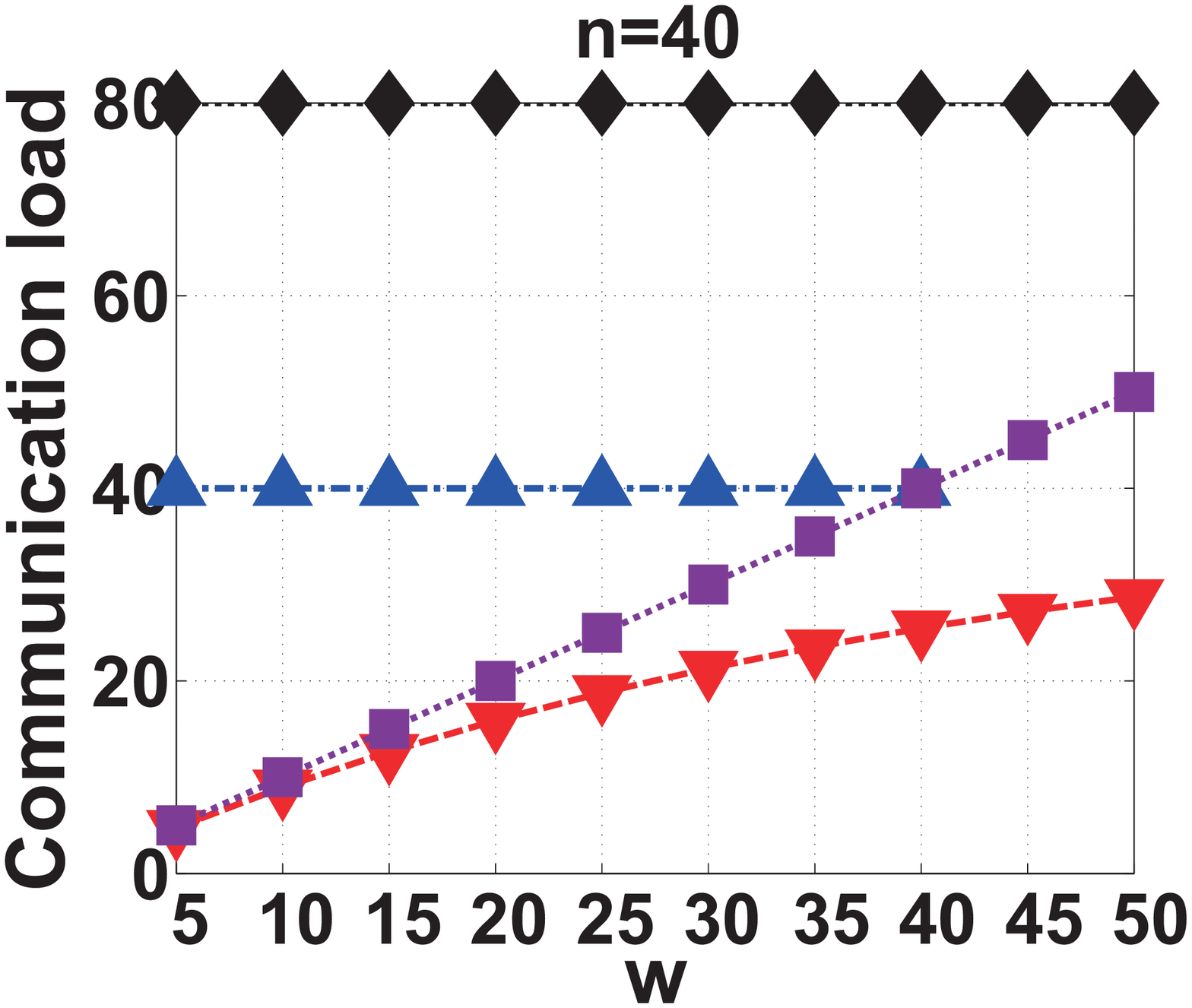}
%\mbox{\footnotesize (a) The private polynomial code.}
\end{minipage}%
\begin{minipage}[t]{0.2\linewidth}
\centering
\includegraphics[width=1.45in]{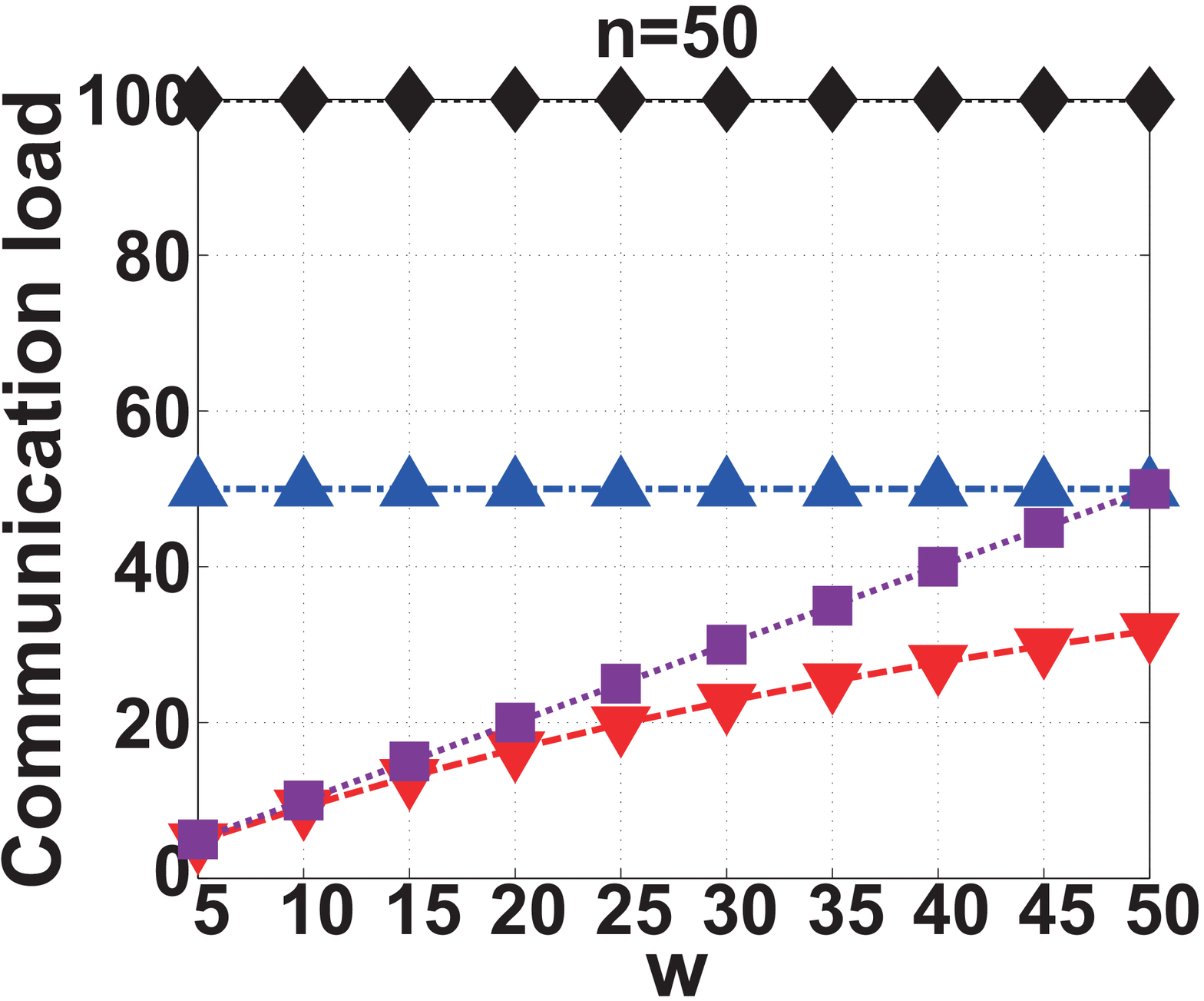}
%\mbox{\footnotesize (a) The private polynomial code.}
\end{minipage}%
\begin{minipage}[t]{0.2\linewidth}
\centering
\includegraphics[width=1.45in]{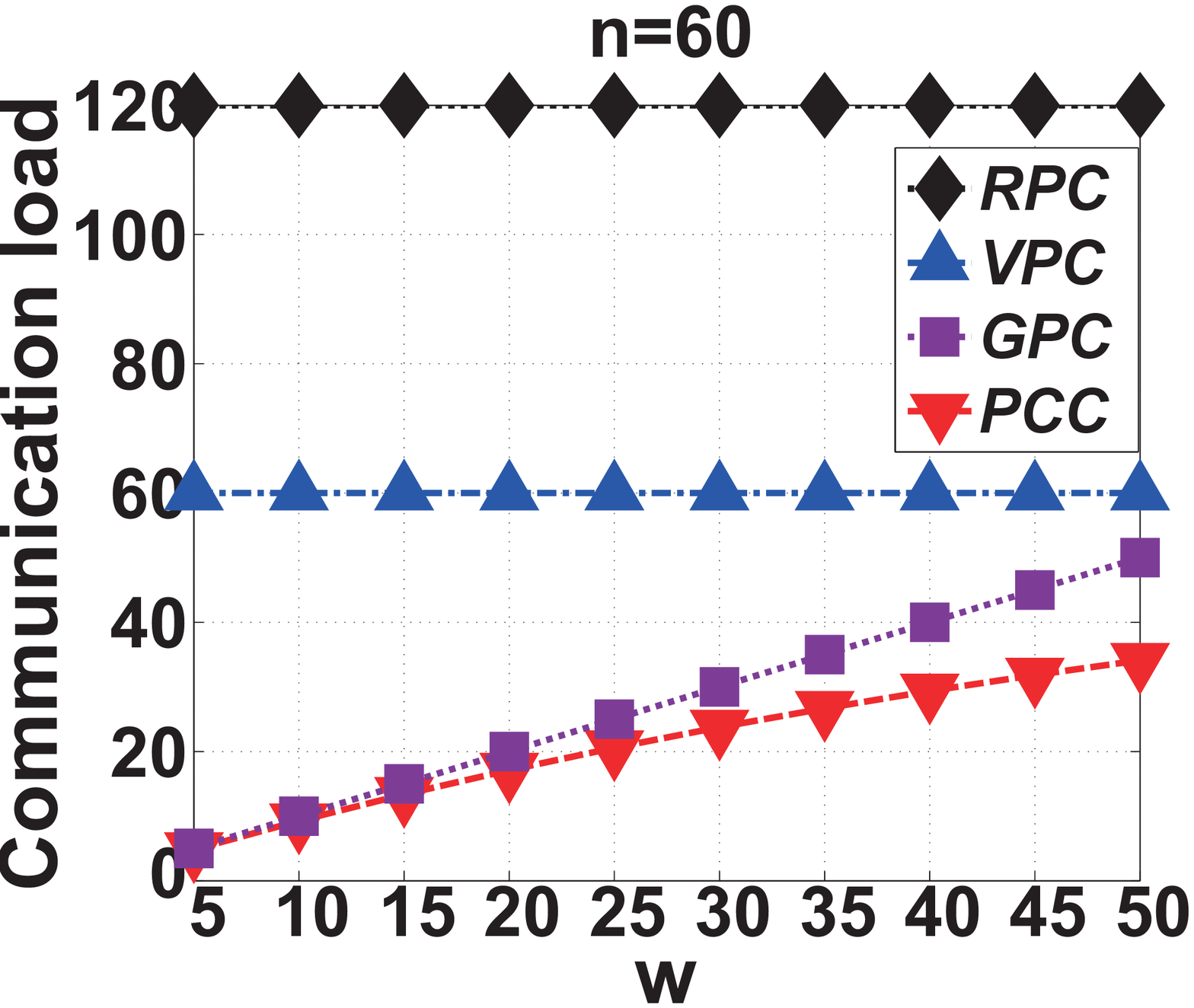}
%\mbox{\footnotesize (a) The private polynomial code.}
\end{minipage}%

\mbox{\footnotesize (a) Each edge device can store at least $w$ blocks.}

\begin{minipage}[t]{0.2\linewidth}
\centering
\includegraphics[width=1.45in]{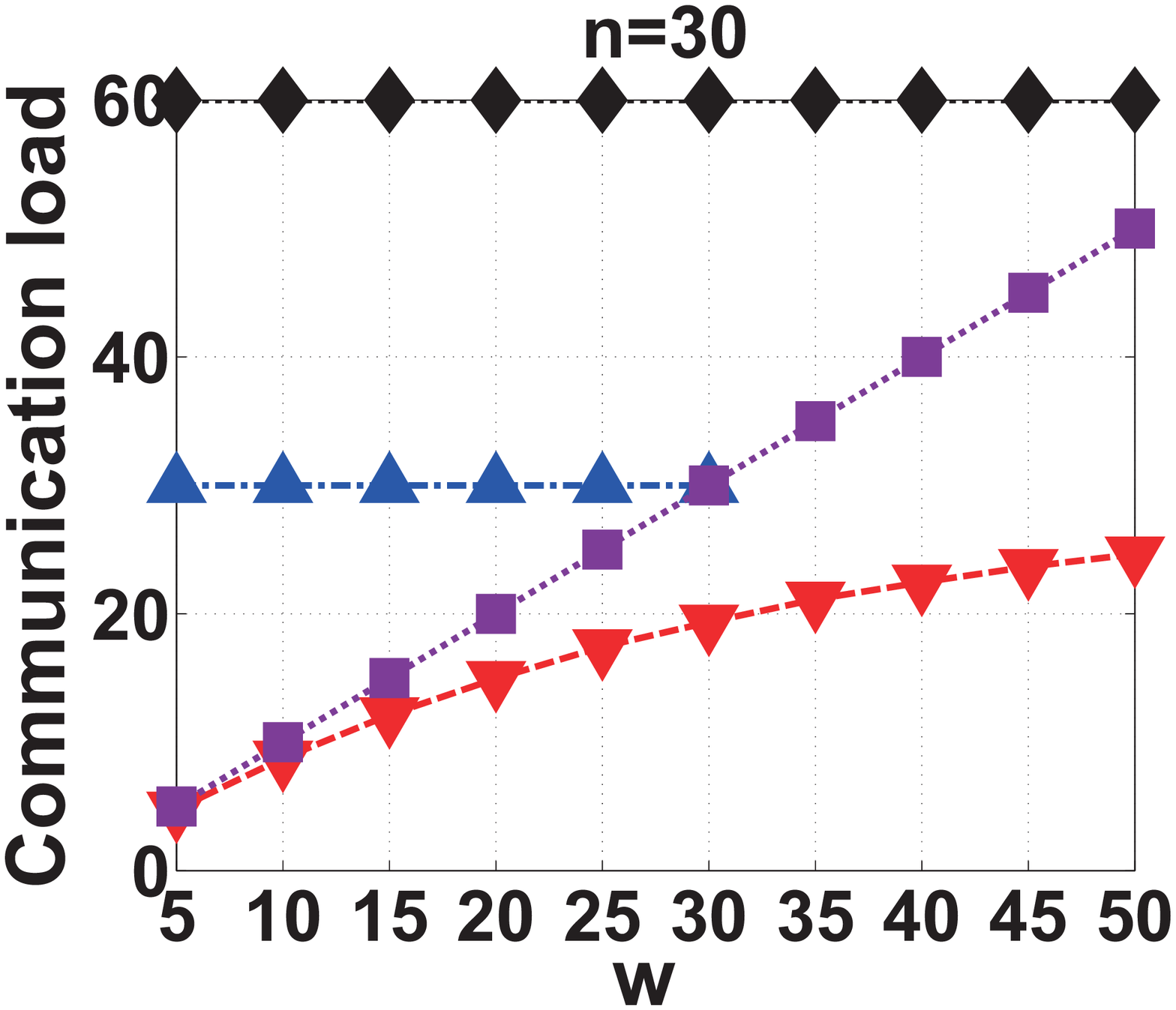}
%\mbox{\footnotesize (a) The private polynomial code.}
\end{minipage}%
\begin{minipage}[t]{0.2\linewidth}
\centering
\includegraphics[width=1.45in]{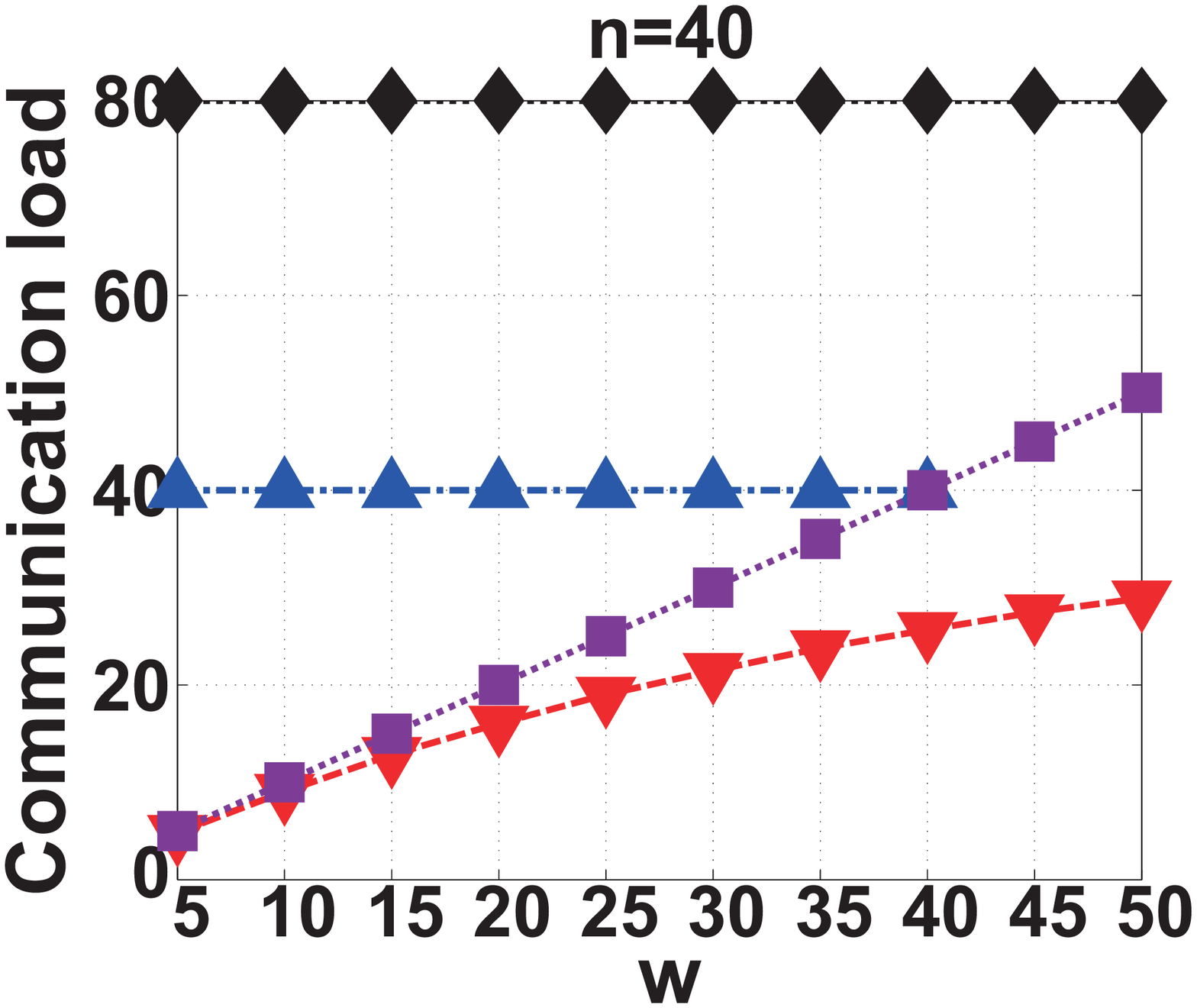}
%\mbox{\footnotesize (a) The private polynomial code.}
\end{minipage}%
\begin{minipage}[t]{0.2\linewidth}
\centering
\includegraphics[width=1.45in]{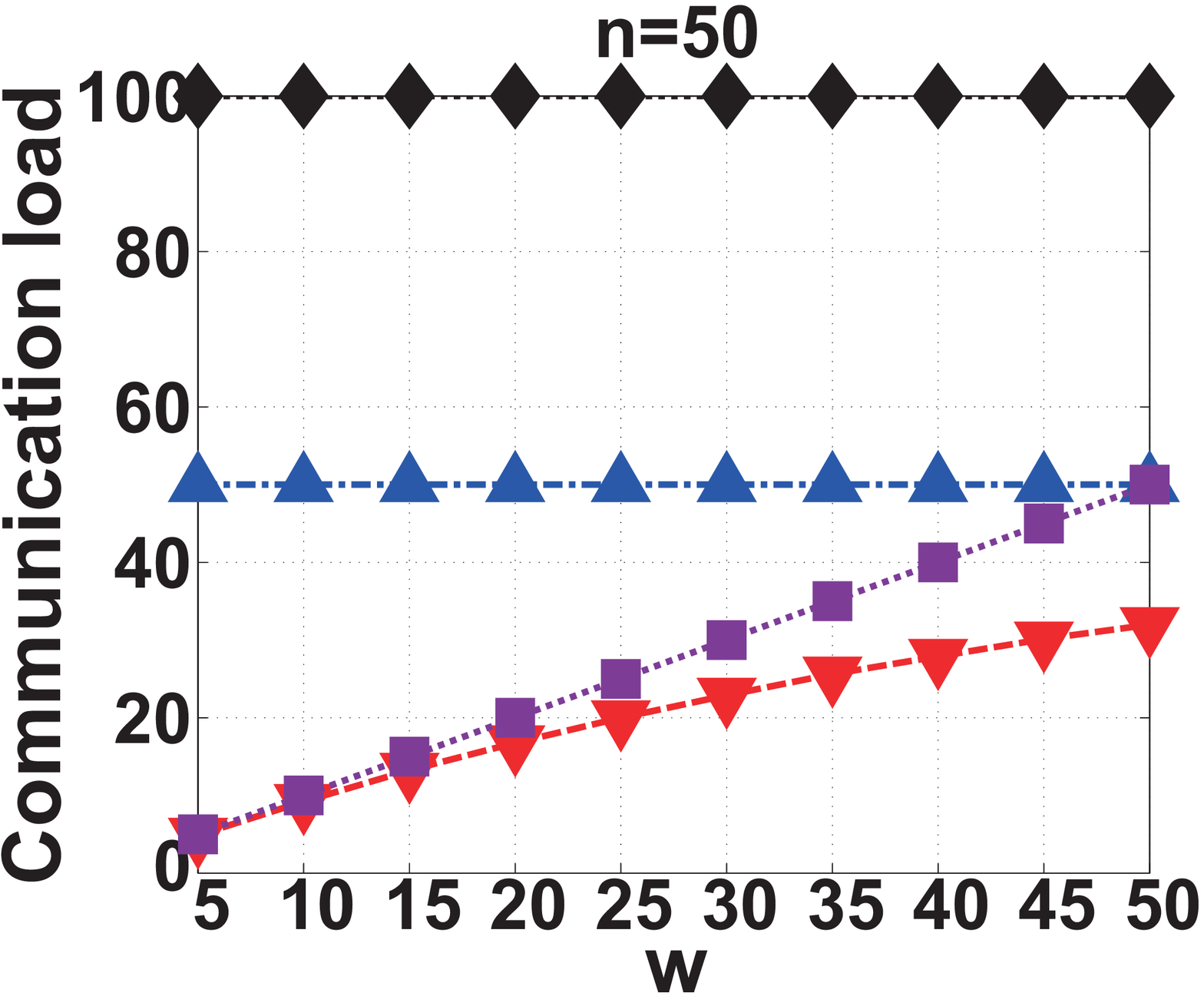}
%\mbox{\footnotesize (a) The private polynomial code.}
\end{minipage}%
\begin{minipage}[t]{0.2\linewidth}
\centering
\includegraphics[width=1.45in]{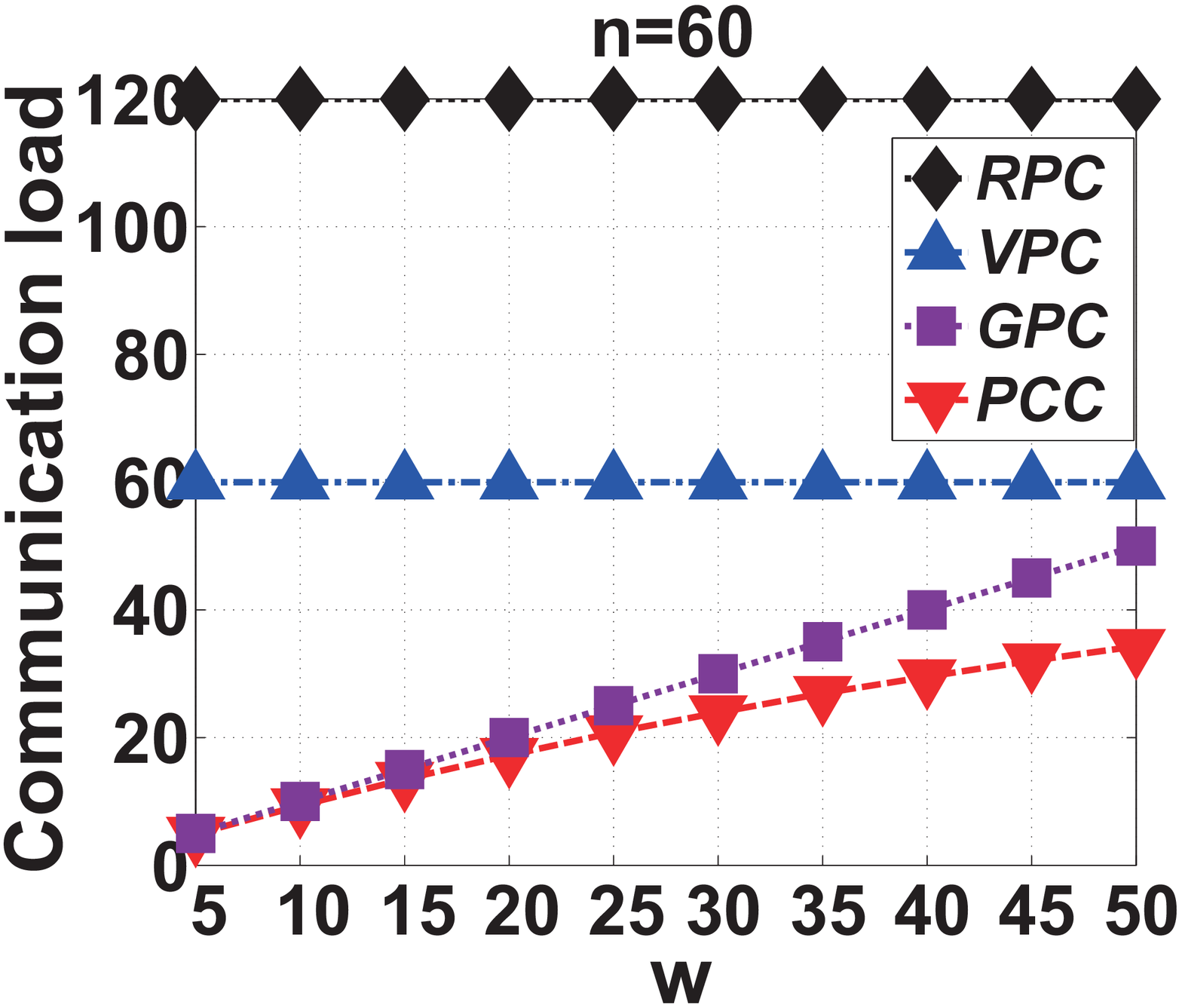}
%\mbox{\footnotesize (a) The private polynomial code.}
\end{minipage}%

\mbox{\footnotesize (b) Each edge device stores $w/2$ blocks at most.}

\begin{minipage}[t]{0.2\linewidth}
\centering
\includegraphics[width=1.45in]{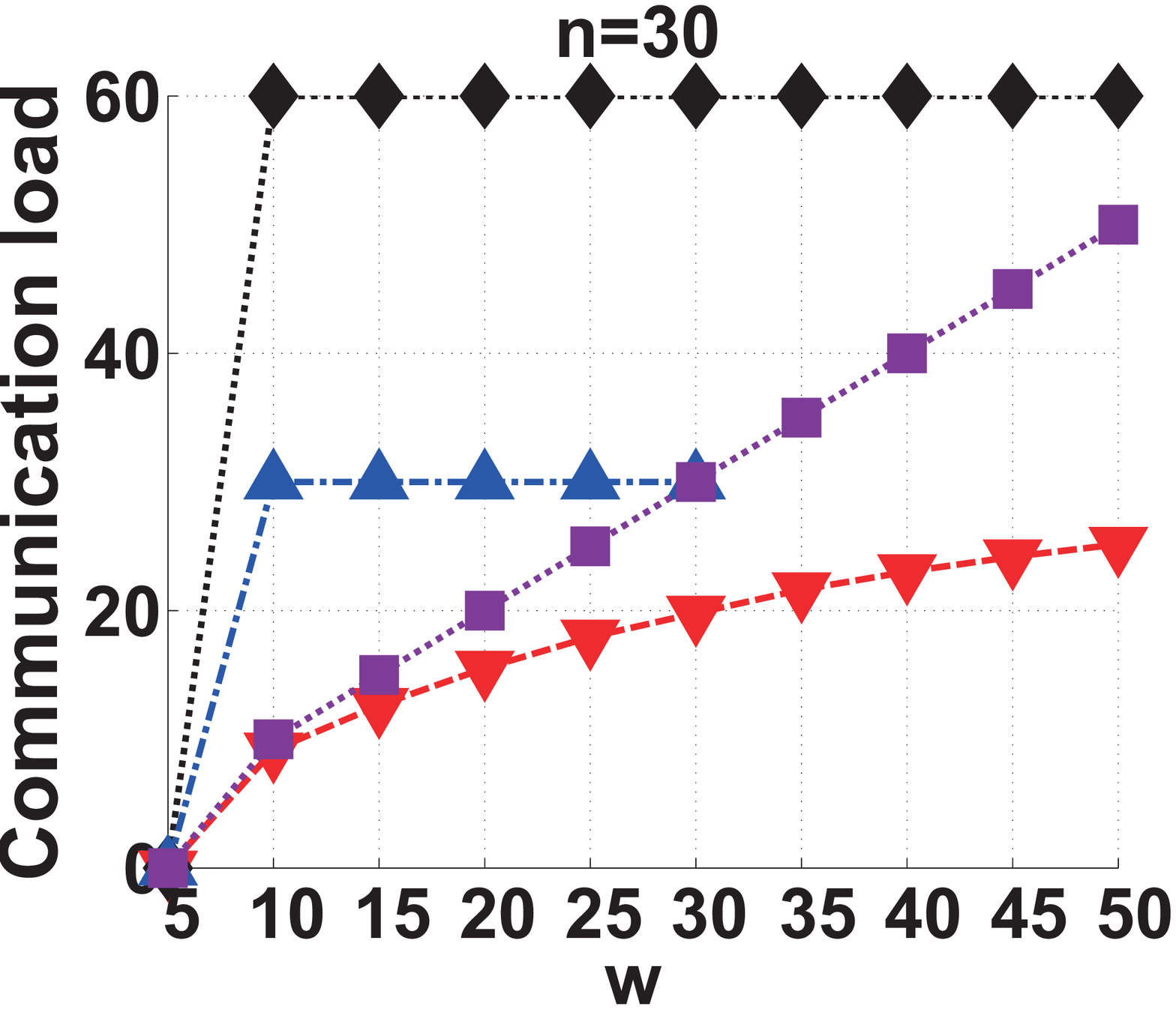}
%\mbox{\footnotesize (a) The private polynomial code.}
\end{minipage}%
\begin{minipage}[t]{0.2\linewidth}
\centering
\includegraphics[width=1.45in]{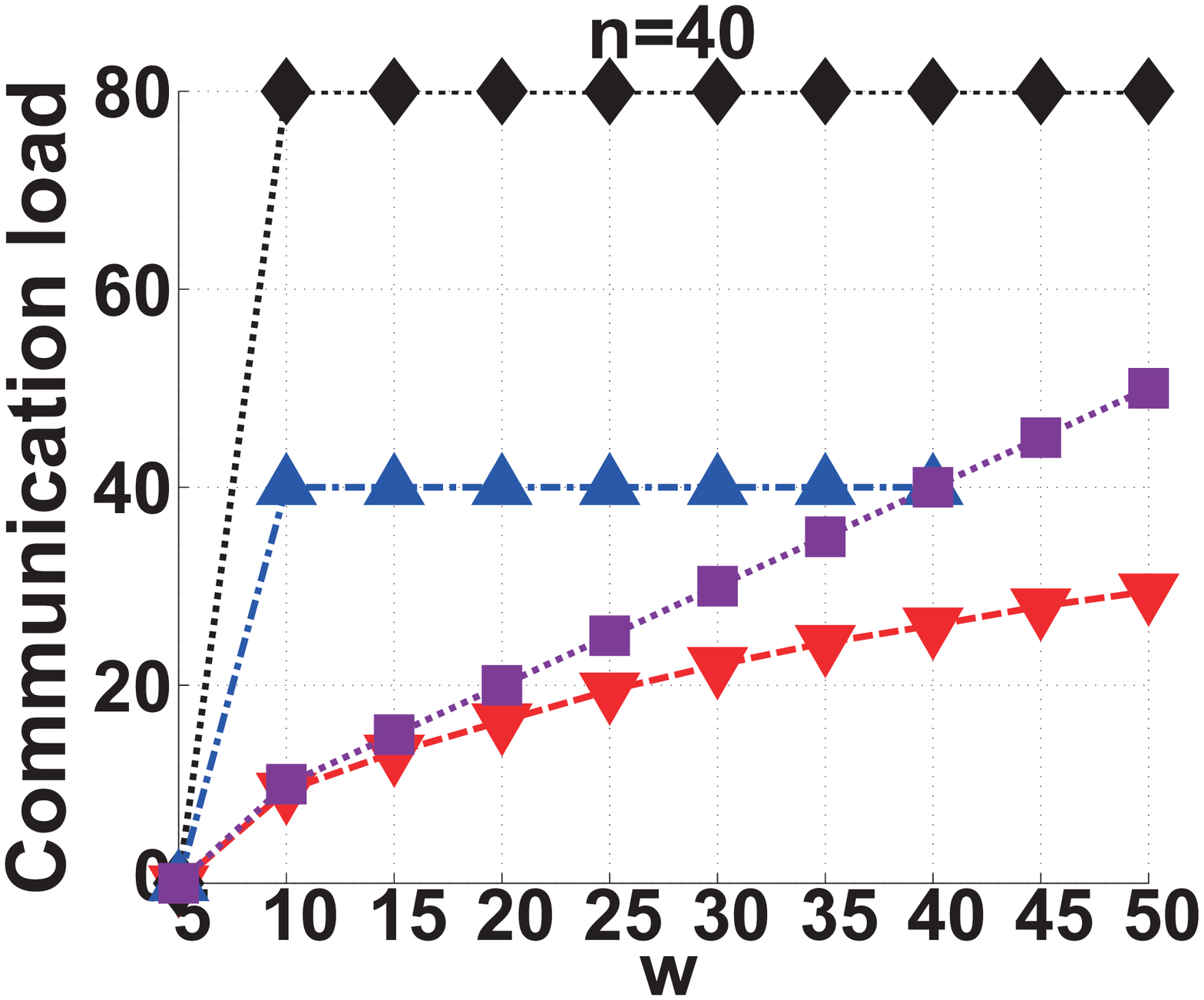}
%\mbox{\footnotesize (a) The private polynomial code.}
\end{minipage}%
\begin{minipage}[t]{0.2\linewidth}
\centering
\includegraphics[width=1.45in]{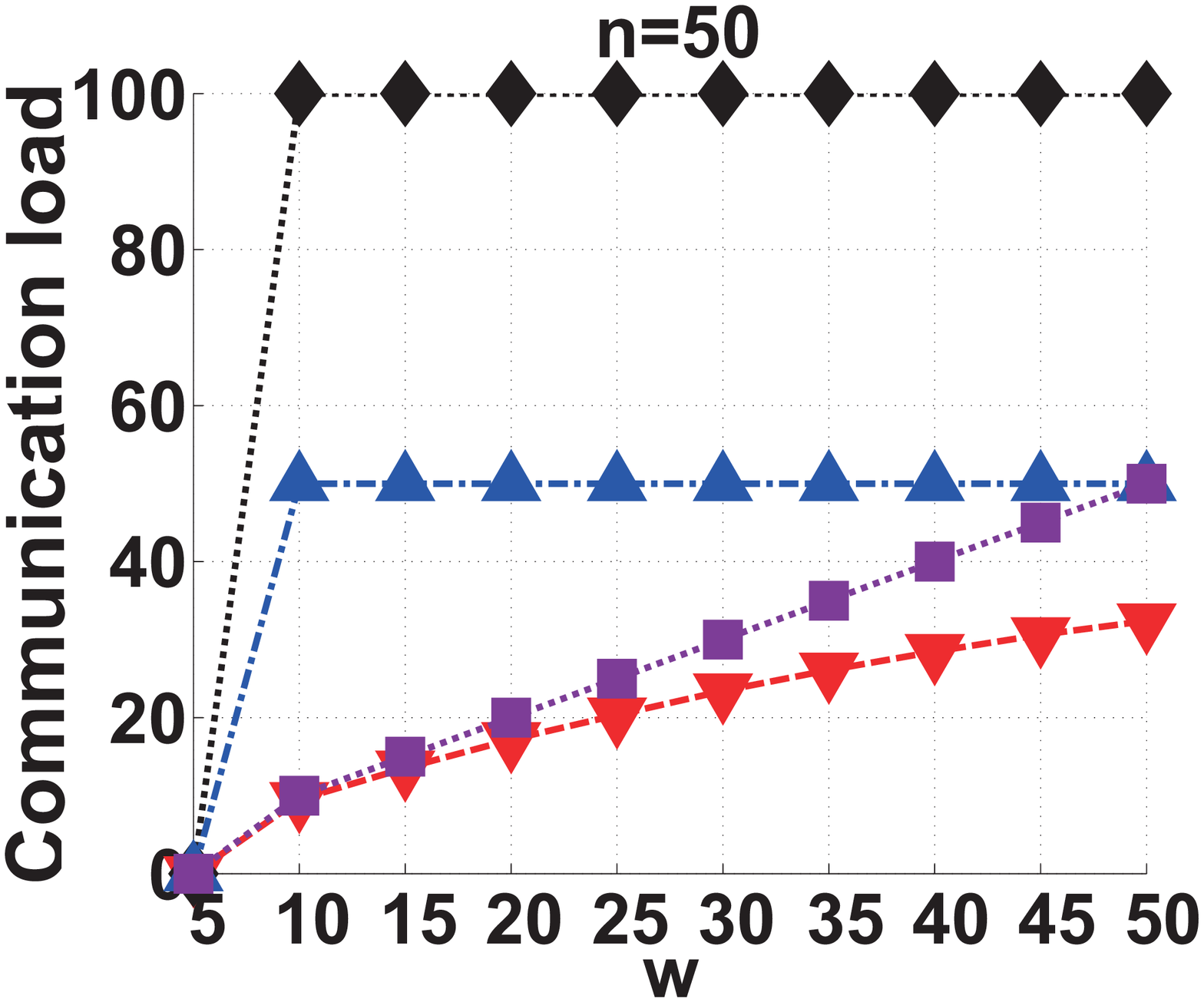}
%\mbox{\footnotesize (a) The private polynomial code.}
\end{minipage}%
\begin{minipage}[t]{0.2\linewidth}
\centering
\includegraphics[width=1.45in]{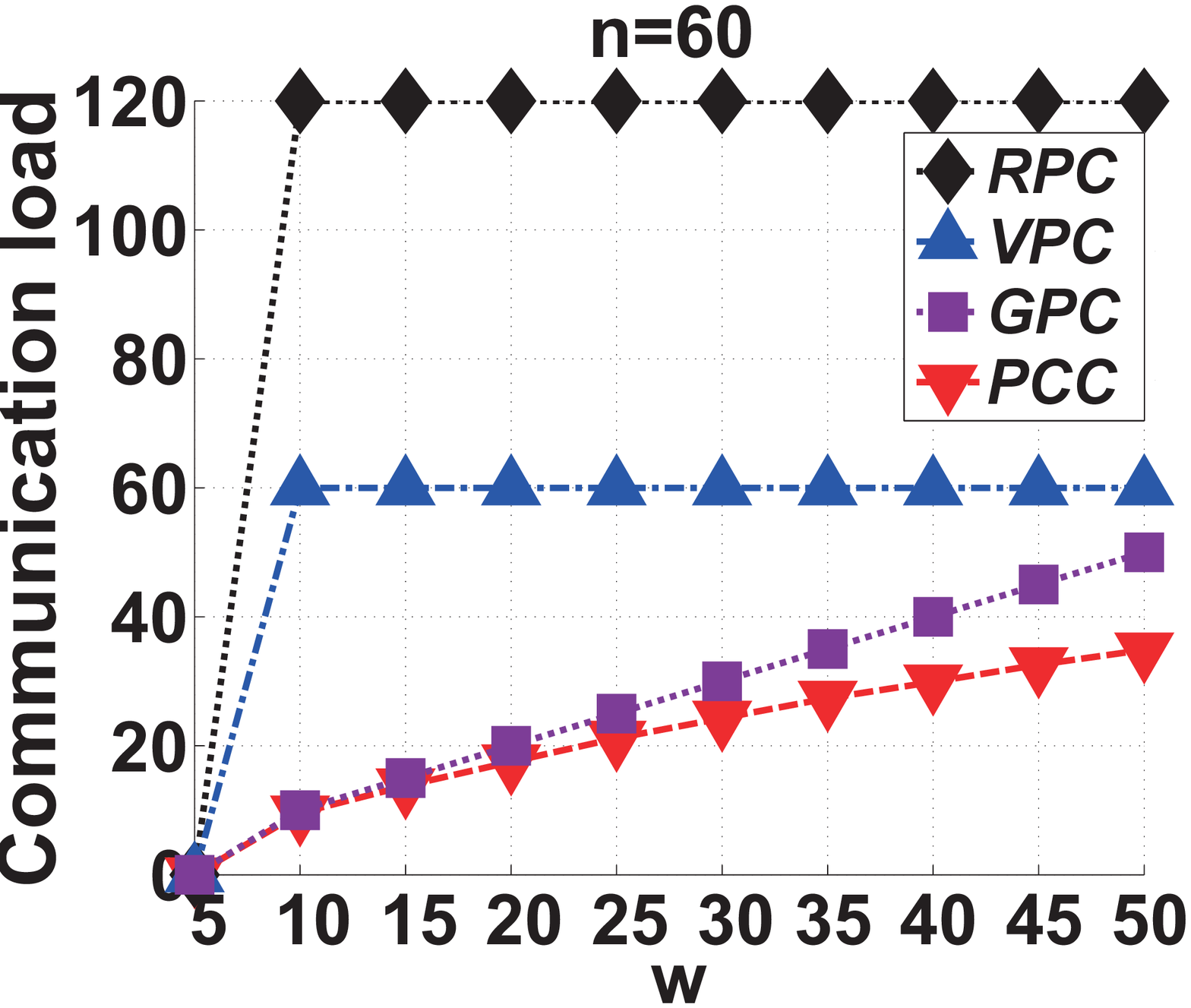}
%\mbox{\footnotesize (a) The private polynomial code.}
\end{minipage}%

\mbox{\footnotesize (c) Each edge device stores $w/4$ blocks at most.}

\begin{minipage}[t]{0.2\linewidth}
\centering
\includegraphics[width=1.45in]{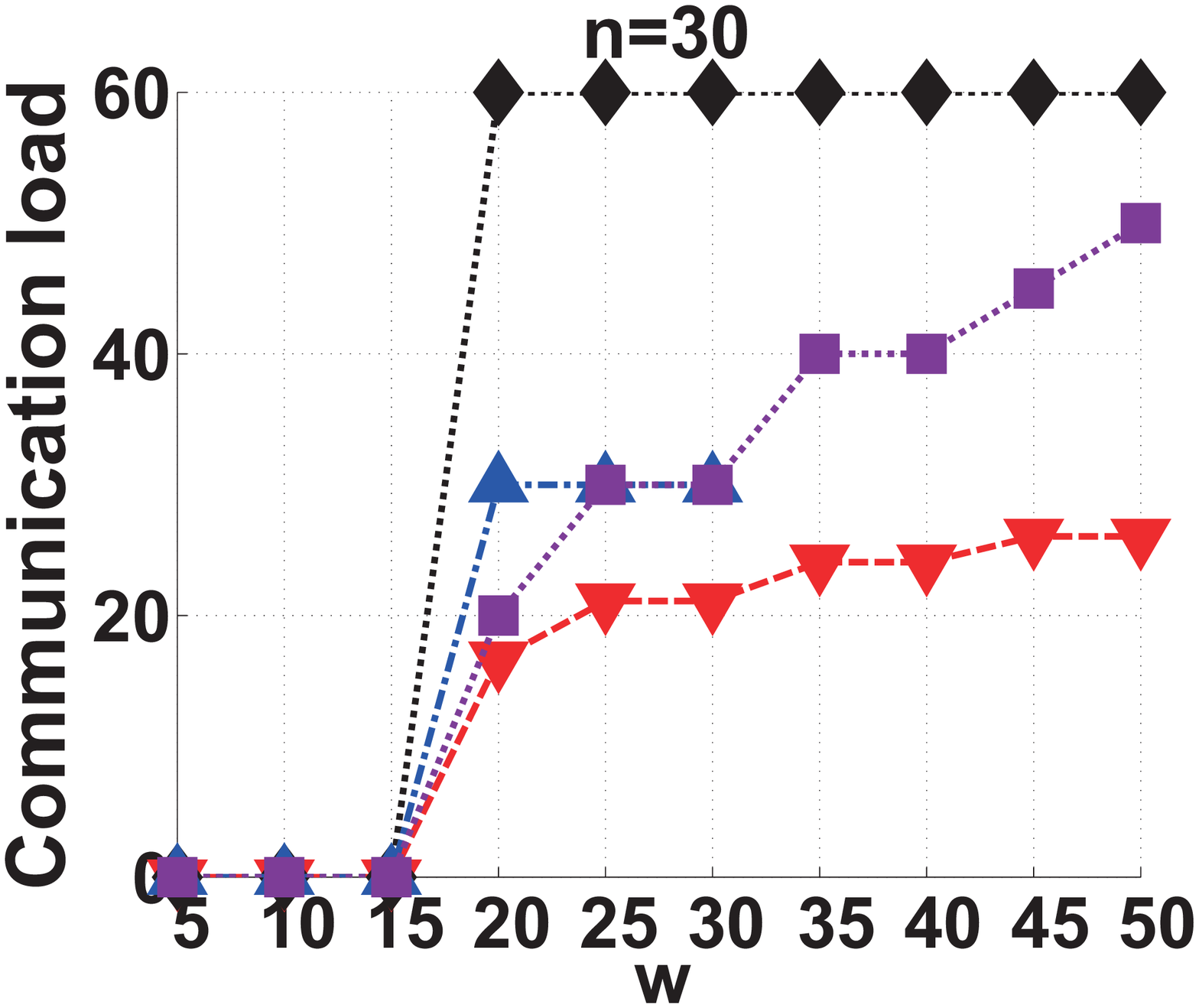}
%\mbox{\footnotesize (a) The private polynomial code.}
\end{minipage}%
\begin{minipage}[t]{0.2\linewidth}
\centering
\includegraphics[width=1.45in]{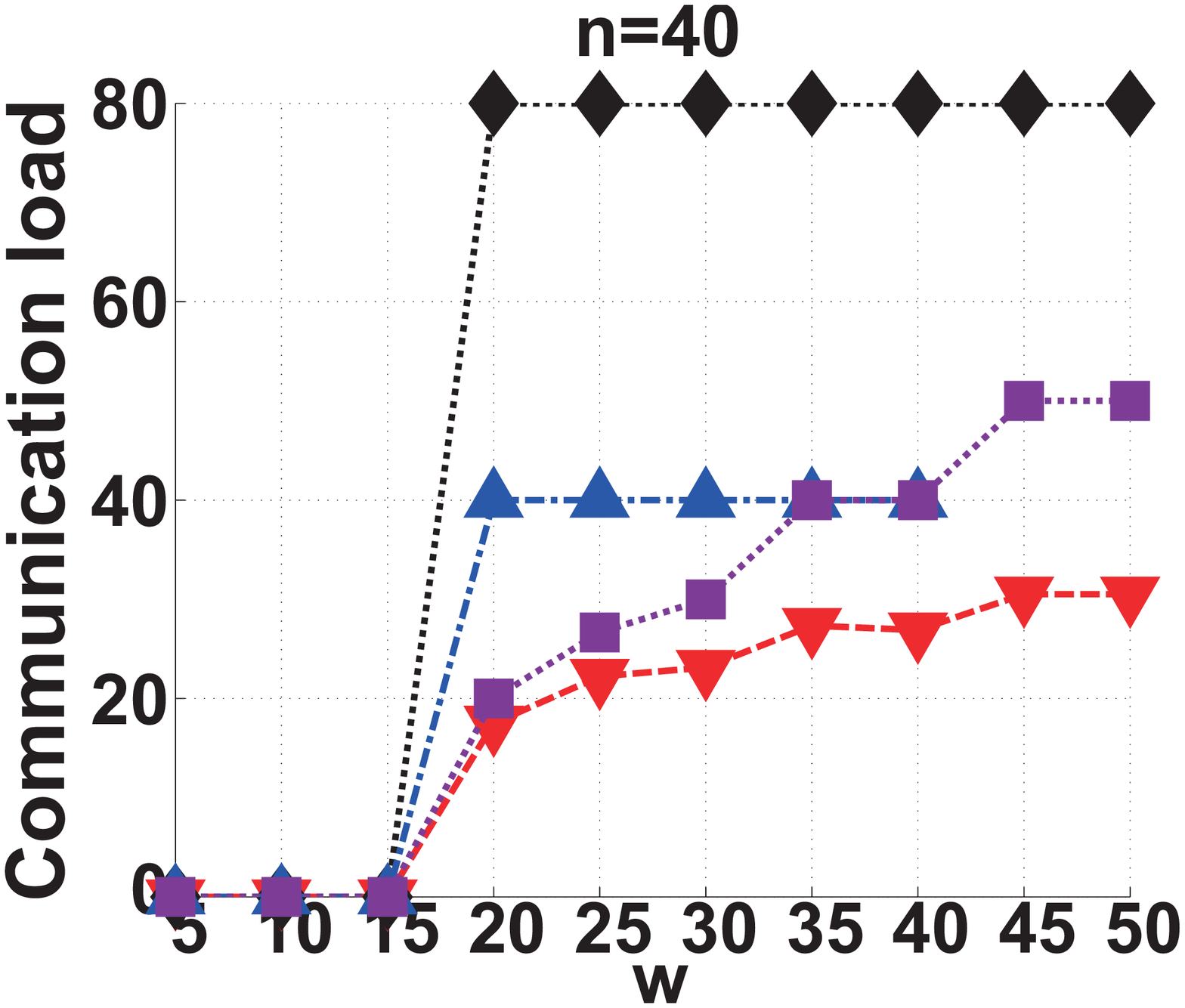}
%\mbox{\footnotesize (a) The private polynomial code.}
\end{minipage}%
\begin{minipage}[t]{0.2\linewidth}
\centering
\includegraphics[width=1.45in]{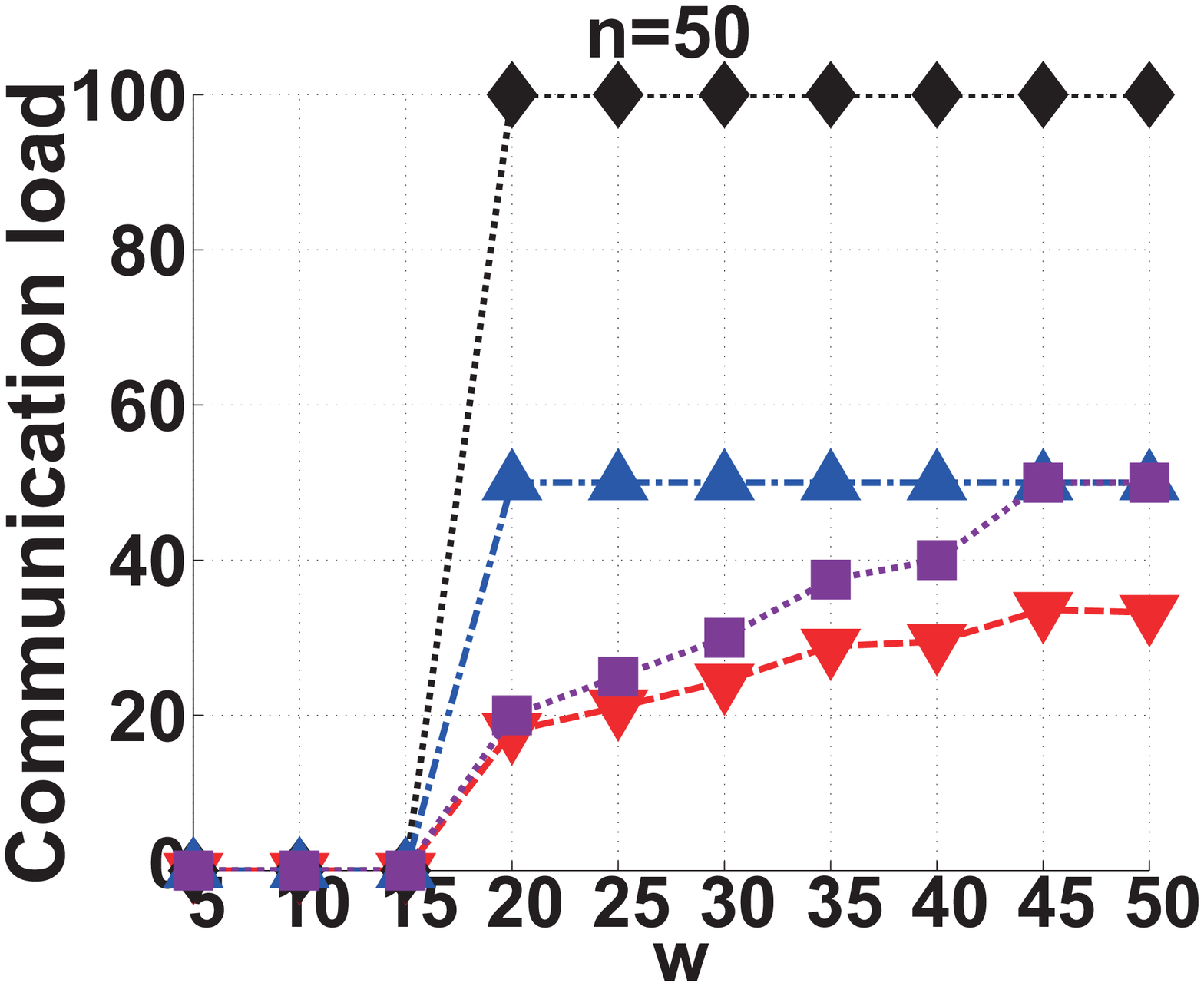}
%\mbox{\footnotesize (a) The private polynomial code.}
\end{minipage}%
\begin{minipage}[t]{0.2\linewidth}
\centering
\includegraphics[width=1.45in]{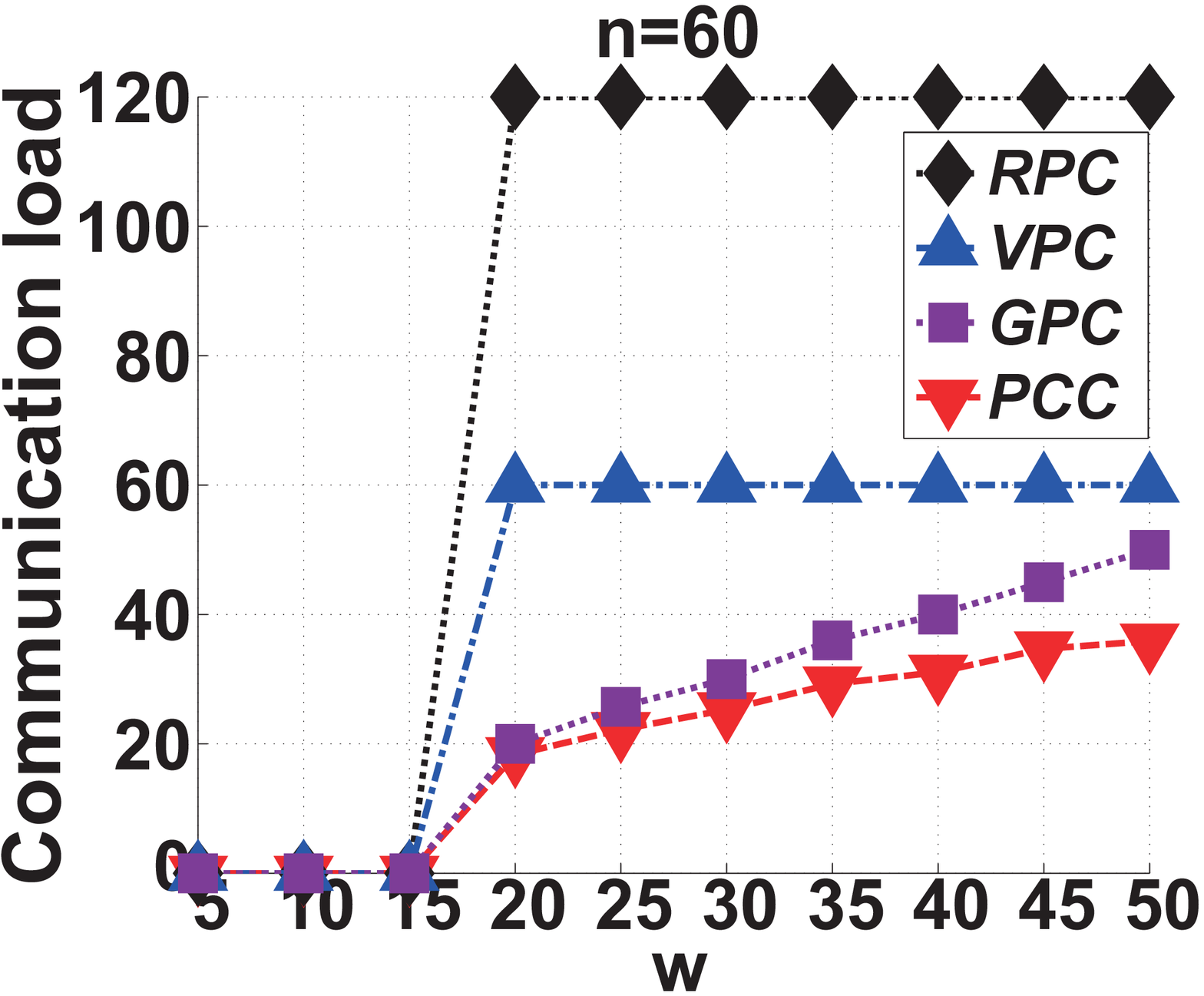}
%\mbox{\footnotesize (a) The private polynomial code.}
\end{minipage}%

\mbox{\footnotesize (d) Each edge device stores $w/8$ blocks at most.}

\caption{Comparison of communication load of different schemes.}
\label{Fig.experiment}
\end{figure*}

In \figref{Fig.experiment}, we show the minimum communication load of each scheme, where the storage limit $t_0$ changes from $w$ to $w/8$. Each scheme may choose different values of $t$ to achieve its minimum communication load.
From a single diagram, it can be observed that the communication load of GPC or PCC increases with the increase of $w$. While that of RPC or VPC is independent of $w$. Moreover we have the following observations: (1) PCC outperforms GPC (in $70\%$ the load), VPC (in $50\%$ the load), RPC (in $25\%$ the load); (2) VPC cannot complete the computing tasks when $n<w$ as shown in \tabref{tab:complex}, which means in terms of the applicability, GPC, RPC and PCC outperform VPC; (3) in \figref{Fig.experiment} (c) and (d), the communication load of each scheme is 0 when $w$ is small. It's because $t_0<2$ in those points. As mentioned in \secref{SA}, we don't discuss the private problem then.
In \figref{Fig.experiment} (a), we find that the communication load of GPC is independent of $n$.
In contrast, the communication loads of RPC and VPC increase linearly with $n$. Meanwhile the increase in PCC is slowly and negligible.
Comparing \figref{Fig.experiment} (a)-(d), we find that with the decrease of $t_0$, the increase of communication load in GPC or PCC is relatively small.

Concluding from the simulations, PCC and GPC can be applied in more cases than VPC. Besides, PCC and GPC always outperform VPC and RPC in terms of communication load. In particular, compared with GPC, VPC and RPC, PCC reduces the total communication load by $30\%$, $50\%$, $75\%$ respectively.

	\section{Conclusion}
	\label{sec.conc}

	In this paper, we study the private matrix multiplication in EC scenario, where edge devices may only store part of the library. We focus on the joint research of storage allocation and computation design in an unified framework. We first give a general storage allocation scheme to distribute the library to edge devices. Then we design two efficient computing schemes to protect the user's privacy with lower communication load. Moreover we give the theoretical analysis on the proposed computing schemes. Finally, we conduct extensive simulation experiments to show the efficiency of the proposed computing schemes. We will consider implementing the proposed schemes in real edge computing systems in future works.

In this paper, we study the PEC problem where edge devices have limited storage and communication resources. Although there are schemes valid for the private problem, but they are either not useful for EC or with high communication load. We focus on the joint research of storage allocation and computation design, and make a tradeoff between the storage and communication resources, under the circumstance of (1) completing the computing tasks, (2) protecting the user's privacy, and (3) decreasing the communication load.
Firstly, we give a general solution to arrange the library for any possible value of storage factor.
Then based on the storage allocation, we design two efficient private computing schemes to protect the user's privacy with lower communication load. Moreover we give the theoretical analysis on the two schemes and show that they allow the tradeoff between the storage and the communication resources.
Finally, we conduct extensive simulation experiments, which demonstrate the effectiveness of the proposed computing schemes. We will consider implement the proposed schemes in real edge computing systems in future works.

There are a few interesting follow-up research directions of this work:
\begin{itemize}
		\item
		\emph{Encoding storage}: In this paper, the storage factor $t$ is a positive integer. However, it can be a fraction too. In this situation, we can split the blocks first and immigrate part of them or their coded results to the edge devices. By encoding the block in the allocation phase, we can make full use of storage resource and even ensure the confidentiality of $\textbf{B}$.
		\item
		\emph{Cooperative attack}: In this paper, the edge devices don't collude with each other. However in a real work environment, the relationship between edge devices may be more complex. We should study a more general case where edge devices may attack cooperatively.
		\item
		\emph{Heterogeneous networks}: We can also consider several heterogeneous networks. In EC, the devices often come from different networks and their costs of storage, communication and computing are different. They may have different
performance to compute subtasks, store data and communicating with each other. In this case, we can complete the private computing with the objective to minimize the total cost.
	\end{itemize}

		\bibliographystyle{IEEEtran}	
		\bibliography{PEC}

\end{document}